\documentclass[journal]{IEEEtran}
\ifCLASSINFOpdf

\else

\fi
\usepackage{mathrsfs}

\usepackage{amsmath}
\usepackage{ntheorem}

\usepackage{makeidx}  
\usepackage{algorithm}
\usepackage{algorithmic}
\usepackage{graphicx}
\usepackage{subfigure}
\usepackage{epstopdf}
\usepackage{bm}
\usepackage{cite}
\usepackage{stfloats}
\usepackage{setspace}

\usepackage{amssymb}
\usepackage{graphicx}
\newtheorem{theorem}{Theorem}
\usepackage{url}

\usepackage{tabularx,booktabs}
\newcolumntype{C}{>{\centering\arraybackslash}X} 
\usepackage{lipsum}

\usepackage{makecell} 

\usepackage{graphicx}
\usepackage{color}

\newtheorem{lem}{Lemma}
\newtheorem{remark}{Remark}

\usepackage{multirow}

\definecolor{dblue}{RGB}{15,89,164}
\usepackage{amsmath,amssymb}

\begin{document}

\title {  Holographic MIMO Multi-Cell Communications }
\author{ Kangda Zhi, Tianyu Yang, Shuangyang Li,  Yi Song, Tuo Wu, and Giuseppe Caire, {\itshape Fellow, IEEE \upshape}		
	\thanks{ Kangda Zhi, Tianyu Yang, Shuangyang Li,  Yi Song,  and Giuseppe Caire are with Communications and Information Theory Group (CommIT), Technische Universit\"{a}t Berlin, 10587 Berlin, Germany (e-mail: \{k.zhi, tianyu.yang, shuangyang.li, yi.song, caire\}@tu-berlin.de).}
	\thanks{Tuo Wu is with School of Electrical and Electronic Engineering, Nanyang Technological University, 639798, Singapore (E-mail: tuo.wu@ntu.edu.sg). }	
	 \vspace{-20pt}
}

\maketitle
\begin{abstract}
	Metamaterial antennas are appealing for next-generation wireless networks due to their simplified hardware and much-reduced size, power, and cost. This paper investigates the holographic multiple-input multiple-output (HMIMO)-aided multi-cell systems with practical per-radio frequency (RF) chain power constraints. With multiple antennas at both base stations (BSs) and users, we design the baseband digital precoder and the tuning response of HMIMO metamaterial elements to maximize the weighted sum user rate. Specifically, under the framework of block coordinate descent (BCD) and weighted minimum mean square error (WMMSE) techniques, we derive the low-complexity closed-form solution for baseband precoder without requiring bisection search and matrix inversion. Then, for the design of   HMIMO   metamaterial elements under binary tuning constraints, we first propose a low-complexity suboptimal algorithm with closed-form solutions by exploiting the hidden convexity (HC) in the quadratic problem and then  further propose an accelerated sphere decoding (SD)-based algorithm which yields global optimal solution in the iteration. For HMIMO metamaterial element design under the Lorentzian-constrained phase  model, we propose a maximization-minorization (MM) algorithm with closed-form solutions at each iteration step. Furthermore, in a simplified multiple-input single-output (MISO) scenario, we derive the scaling law of downlink single-to-noise (SNR) for HMIMO with binary and Lorentzian tuning constraints and theoretically compare it with conventional fully digital/hybrid arrays.  Simulation results demonstrate the effectiveness of our algorithms compared to benchmarks and the benefits of  HMIMO  compared to conventional arrays.
\end{abstract}


\IEEEpeerreviewmaketitle

\vspace{-10pt}
\section{Introduction} \label{section0}

The six-generation (6G) networks require order-of-magnitude higher spectral efficiency than the current fifth-generation networks so that the solid foundation for many fancying commercial applications can be built, such as immersive extended reality  and remote multi-sensory telepresence. To this end, numerous enabling technologies have been proposed and studied in recent years, such as extremely large-scale multiple-input multiple-output (MIMO)\cite{near2025review}, cell-free \cite{shi2024ris}, and  millimeter wave (mmWave) communications\cite{fang2021hybrid}.

Meanwhile, future wireless networks are envisaged to be cost-efficient, sustainable, and environmentally friendly. This motivates a huge amount of research on metamaterial antennas. Compared to conventional antennas, metamaterial antennas consist of sub-wavelength radiators excited by waveguides or cavities with very low power consumption, hardware complexity, and costs\cite{shlezinger2019dynamic,gao2023programmable,gao2024terahertz}. As embodiments of this technology, reconfigurable intelligent surface (RIS) \cite{chen2022active,zhiTwotimescale2022} and holographic MIMO (HMIMO) \cite{huang2020holographic,gong2023holographic,deng2022reconfigurable,you2022energy,zhang2022beam,xu2024near} have recently been proposed and studied.  In this work, we focus primarily on active HMIMO, which operates as a transceiver technology.

HMIMO could be implemented by packing tunable metamaterial radiators with   element size and  spacing much smaller than a half wavelength (practically usually on the order of $\frac{1}{10}$ - $\frac{1}{5}$ wavelength \cite{smith2017analysis}) into compact arrays, and therefore it is possible to achieve a nearly continuous planar aperture\cite{Pizzo2020HMIMO}. Imitating the concept of optical holography, HMIMO could flexibly manipulate the electromagnetic (EM) wave and achieve high-precision beamforming based on the holographic principle\cite{deng2023reconfigurable}. Through patterning metamaterial elements onto the upper conductor of microstrip transmission lines and  exciting them by the feed wave along the waveguide\cite{smith2017analysis,boyarsky2017synthetic}, waveguide-fed HMIMO opens up a new kind of array design paradigm with very low hardware cost and  high power efficiency, which is appealing for supporting future cost-efficient wireless networks.

Given these new features,  waveguide-fed metamaterial HMIMO has  recently attracted significant  research interest. Specifically, the waveguide-excited linear metamaterial array was modeled in \cite{smith2017analysis} by polarizable dipoles which couple the waveguide mode to radiation modes, and  the radiation pattern was analyzed and simulated via full-wave simulation software. Based on this model, massive MIMO systems supported by waveguide-excited HMIMO were studied in \cite{shlezinger2019dynamic} and further combined with wideband orthogonal frequency division multiplexing (OFDM) transmissions in \cite{wang2020dynamic}.
In \cite{zhang2022beam}, the beam focusing performance of HMIMO was studied in near fields under Lorentzian phase constraints.  The advantages of HMIMO on energy efficiency (EE) performance compared to conventional arrays were elaborated in \cite{you2022energy} under different availability of channel state information. The design of HMIMO based on the low-complexity holography principle was conducted in \cite{deng2022reconfigurable,zhang2022holographic} and the hardware prototype was developed for experimental measurements\cite{deng2023reconfigurable}.

However, all of the above-mentioned work focused on  single-cell networks. In  practical cellular systems, due to the reuse of the limited frequency resource, interference from nearby cells is inevitable. Without an effective treatment,  inter-cell interference will significantly degrade the overall network performance, with a particularly unfavorable impact on the service experience of cell-edge users. For  HMIMO-enabled networks with the new metamaterial antenna architecture,  both the multi-cell system performance and the design of effective signal processing algorithm are fully unexplored, which motivates our work. 

Besides, compared with the sum power constraint over all antenna ports which is commonly adopted in  designing digital precoding, it is more realistic to consider the individual power constraint with respect to each radio-frequency (RF) chain. This is because  in  practice, each RF chain is equipped with its own power amplifier and  therefore,  is individually  power-limited\cite{dsouza2018hybrid,zhao2023rethinking,hu2023single}. Meanwhile,  the usable RF output power of each RF chain amplifier is constrained by its linearity range, which in turn determines the consumed DC power\footnote{This depends on the amplifier technology, quality, and cost (e.g., see \cite{wang2020power}), and its discussion goes beyond the scope of this paper.}. Hence, enforcing a per-RF chain power constraint is physically meaningful since this constraint determines the required linear range and hence eventually the actual DC power spent in the amplification hardware.  However, to the best   of our knowledge, this practical constraint has not been considered in designing HMIMO.

Furthermore, two fundamental questions about  HMIMO  remain unanswered from theoretical perspectives. First, when deploying HMIMO with various types of metamaterial elements, how does the received  signal-to-noise (SNR)  scale  with the number of tunable elements? Secondly,  given the same total power budget,
can HMIMO achieve better performance than conventional fully digital/hybrid array?  Investigating these questions is crucial for us  to deeply understand the   properties of  HMIMO and the working mechanism behind it.

Motivated by the above research gaps, in this paper, we study the  multi-cell networks aided by HMIMO, where both the base stations (BSs) and users are equipped with multiple antennas. We design baseband precoder under the per-RF chain power constraints and design HMIMO tunable elements under binary/Lorentzian phase constraints, respectively. We also leverage a simplified multiple-input single-output (MISO) scenario to  draw theoretical insights on the properties of HMIMO compared with conventional arrays. Our main contributions are summarized as follows.
 		\begin{itemize}
\item We formulate the weighted sum user rate maximization problem for HMIMO-enabled downlink multi-cell multiple-user networks, under the per-RF chain power constraint for digital precoder and binary/Lorentzian constraints for HMIMO tunable elements.

\item By exploring the problem structure, we propose closed-form solutions for digital precoder design under the per-RF chain power constraint, without  bisection search and matrix inversion operations so that   very low computational complexities are required.

\item To configure HMIMO elements under binary   constraints, we first propose a closed-form low-complexity sub-optimal solution by exploiting the hidden convexity (HC) feature in the reformulated quadratic problem. Next, to further improve the performance and help evaluate the effectiveness of the HC algorithm, we propose an accelerated sphere decoding (SD) algorithm to obtain the globally optimal solution in each iteration.

\item  To configure HMIMO elements under grayscale constraints,  following the Lorentzian-constrained phase model, we propose a low-complexity maximization-minorization (MM)  algorithm with closed-form solutions in each iteration.

\item In MISO scenarios,  we  derive the SNR  expressions for the digital array, hybrid digital/analog (D/A) array, and HMIMO with binary and Lorentzian constraints, respectively, under the same total power budget. We prove that  SNRs of HMIMO have the same scaling law as conventional arrays but with lower coefficients.

\item Based on numerical results, we  demonstrate the effectiveness of the proposed algorithms and validate the correctness of our analytical conclusions. Besides, we explore the great potential of a new center-feeding-type HMIMO to improve the quality of analog beamforming.
\end{itemize}

The rest of this paper is organized as follows. Section \ref{section1} formulates the  multi-cell  optimization problem. Sections \ref{section2} and \ref{section3}  respectively design the baseband precoder and  HMIMO tunable response.  Section \ref{section4} derives and analyzes the SNR scaling law of HMIMO in the MISO case. Section \ref{section5} provides numerical results and Section \ref{section6} concludes this work.

\vspace{-15pt}
\section{System Model}\label{section1}

\begin{figure}[t]
	\setlength{\abovecaptionskip}{-5pt}
\setlength{\belowcaptionskip}{-15pt}
	\centering
	\includegraphics[width= 0.45\textwidth]{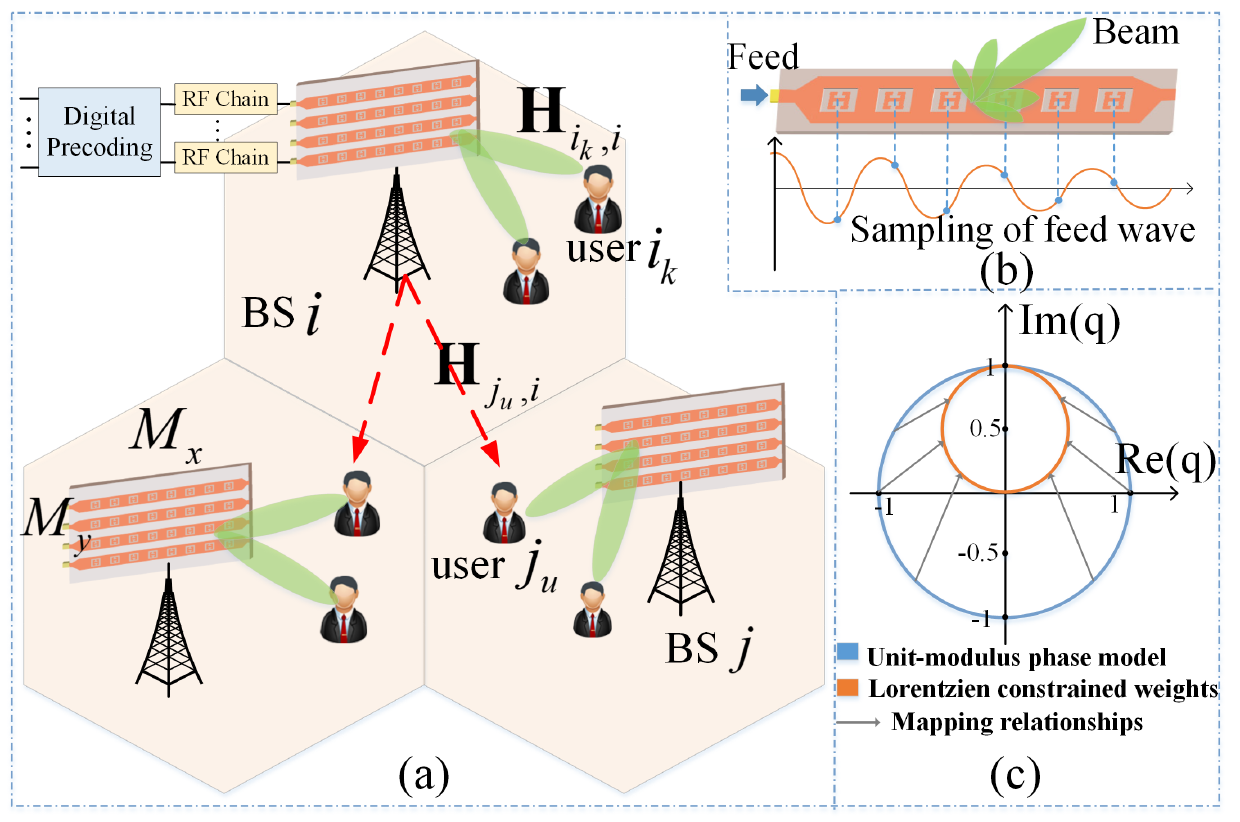}
	\DeclareGraphicsExtensions.
	\caption{Illustrations of (a)  multi-cell HMIMO networks; (b) the excitation of metamaterial elements by the feed wave; (c) constraints of tunable weights.}
	\vspace{-10pt}
	\label{figure2}
\end{figure}

As shown in Fig. \ref{figure2}. (a), we consider an HMIMO-enabled downlink MIMO network with  $L$ cells. In each cell $i$,  $U_i$ users are served by a BS equipped with HMIMO. Following hardware implementation in \cite{deng2023reconfigurable}, each HMIMO is constructed by deploying $M_y$ microstrips on the array where each microstrip is connected to a single RF chain that feeds the amplified signal into the waveguide, radiating through the $M_x$ metamaterial elements\cite{smith2017analysis}, as illustrated in  Fig. \ref{figure2}. (b). Accordingly, the total number of radiating elements on each HMIMO is $M=M_xM_y$. Besides, each user is equipped with  a conventional array of $N$ antennas and requests $d$ data streams. Thus, we assume $M\geq M_y\geq U_id$ and $N\geq d$.

Define $i_k$  as the  $k$-th user in the $i$-th cell, served by BS  $i$. Then, the signal transmitted by the  $i$-th BS can be given by
\begin{align}
	{\mathbf{x}_i} = \sum\nolimits_{k = 1}^{{U_i}} {{\mathbf{Q}_i}{\mathbf{T}_i}{\mathbf{W}_{{i_k}}}{\mathbf{s}_{{i_k}}}} 
\end{align}
where $\mathbf{W}_{i_k} \in \mathbb{C}^{M_y \times d}$  is the digital beamforming that BS $i$  uses to transmit the signal $\mathbf{s}_{i_k}  \in \mathbb{C}^{d \times 1} $  to  $k$-th user in its cell, $k\in  \{1, \cdots, U_i  \}$, and $\mathbb{E}\left\{ {{\mathbf{s}_{{i_k}}}\mathbf{s}_{{i_k}}^H} \right\} = {\mathbf{I}_d}$. Matrix $ {\mathbf{T}_i}= \operatorname{blkdiag} \{\mathbf{t}_{i, 1}^{\circ}, \mathbf{t}_{i, 2}^{\circ}, \ldots, \mathbf{t}_{i, M_y}^{\circ} \}  \in {^{M \times {M_y}}} $  characterizes
propagation and attenuation of the reference wave along the waveguides on BS $i$, where
 $ \mathbf{t}_{i, m_y}^o \in \mathbb{C}^{M_x \times 1} $, $ 1 \leq m_y \leq M_y $, collects the coefficients of the $m_y$-th microstrip on BS $i$. Considering the  conventional  hardware structure where  signal from the RF chain is fed into the microstrip from the left side\footnote{In simulations, we will explore the great potential of a novel HMIMO structure where the signal is fed into the microstrip from the center.}, the $m_x$-th element  of $\mathbf{t}_{i, m_y}^o $ is modeled by\cite{smith2017analysis,zhang2022beam}
\begin{align}\label{notationsss}
 t_{i, m_y, m_x} \!  \triangleq  \!  [\mathbf{t}_{i, m_y}^{\circ}]_{m_x} \! \! =\!    \exp   \!    \left\{  \!   { - {\rho _{i,{m_y},{m_x}}}     \!  \!   \left( {{\alpha _{i,{m_y}}} + j{\beta _{i,{m_y}}}}   \!   \right)}     \!    \right\} \! \!   ,
\end{align}
where
$ {\rho _{i,{m_y},{m_x}}} = {m_x}{\Delta _{i,{m_y}}} $  denotes the distance from the  feeding port located on the left  of the $m_y$ microstrip  of BS $i$  to its ${m_x}$-th element.
Hardware-dependent parameters ${\Delta _{i,{m_y}}}$, ${\alpha _{i,{m_y}}}$, and ${\beta _{i,{m_y}}}$ represent the element spacing,  waveguide attenuation coefficient, and wavenumber of the  ${m_y}$-th microstrip at the  $i$-th BS, respectively.
${\mathbf{Q}_i} \triangleq \operatorname{diag}\left\{\mathbf{q}_i\right\}  \in {\mathbb{C}^{M \times M}}$ is a diagonal matrix representing the tunable response of $M$ metamaterial elements on HMIMO at cell $i$, and
$ q_{i, m_y, m_x} \triangleq [\mathbf{q}_i ]_{\left(m_y-1\right) M_x+m_x} $
denotes the tunable response of the ${m_x}$-th metamaterial element on the  ${m_y}$-th microscrip of BS $i$.

 The  resonance frequency of  each metamaterial element on the microstrip can be adjusted by varying either the geometry or the local dielectric environment\cite{smith2017analysis}. Specifically, there could be two styles to tune the individual  metamaterial element: binary and grayscale\cite{boyarsky2017synthetic}. Binary tuning enables elements to be tuned from a radiating (on) state to a non-radiating (off) state, for example, with simple and low-cost circuits constructed by PIN diodes\cite{deng2023reconfigurable}. Grayscale tuning enables a shift in the resonance to be introduced, which can provide finer control of the magnitude and phase of each radiating element, for example, with liquid crystal or varactors. Specifically, for the binary-amplitude tuning case with low hardware complexities, the response of each HMIMO antenna is constricted by $ {q_{i,{m_y},{m_x}}} \in {\mathcal{Q}^b} \triangleq \varpi \cdot \left\{ {0,1} \right\} $,
where $ \varpi \in  \mathbb{R}^+  $ is a constant depending on features of metamaterial elements.
For the grayscale tuning case, according to the polarizabilities distribution of Lorentzian resonance of metamaterial elements,  we have \cite{smith2017analysis}
\begin{align}\label{gray_cons}
	& q_{i,{m_y},{m_x}}        \in {\mathcal{Q}^g} \triangleq  \left\{    \left.      (j + {e^{j\phi }})/2    \;  \right|   \phi  \in \left[ {0,2\pi } \right] \right\},
\end{align}
leading to coupled amplitude and phase and reduced  adjustable freedom. As illustrated in  Fig. \ref{figure2}. (c), it can be seen that the adjustable range of $q_{i,{m_y},{m_x}}$ in the complex plane is  a circle    centered at $j/2$ with unit diameter,  which is clearly different from the conventional/ideal unit-modulus constraints.

Denote ${\mathbf{H}_{i_k,j}} \in {\mathbb{C}^{N \times M}}$  as the downlink channel from the BS  $j$ to the  $k$-th user in the cell $i$.  Then,  the  signal received by user ${i_k}$  can be given by
\begin{align}\label{received_signal}
	\begin{array}{l}
		  {\mathbf{y}_{{i_k}}} = {\mathbf{H}_{{i_k},i}}{\mathbf{x}_i} + \sum\nolimits_{j = 1,j \ne i}^L {{\mathbf{H}_{{i_k},j}}{\mathbf{x}_j}}  + {\mathbf{n}_{{i_k}}}  \\
		   =\! \underbrace {{\mathbf{H}_{{i_k},i}}{\mathbf{Q}_i}{\mathbf{T}_i}{\mathbf{W}_{{i_k}}}{\mathbf{s}_{{i_k}}}}_{\text{ desired signal}} \!+\!  \underbrace {\sum\nolimits_{u = 1,u \ne k}^{{U_i}} \! {{\mathbf{H}_{{i_k},i}}{\mathbf{Q}_i}{\mathbf{T}_i}{\mathbf{W}_{{i_u}}}{\mathbf{s}_{{i_u}}}} }_{ \text{intracell interference}}  \\
		   + \underbrace {\sum\nolimits_{j = 1,j \ne i}^L {\sum\nolimits_{u = 1}^{{U_j}} {{\mathbf{H}_{{i_k},j}}{\mathbf{Q}_j}{\mathbf{T}_j}{\mathbf{W}_{{j_u}}}{\mathbf{s}_{{j_u}}}} }  + {\mathbf{n}_{{i_k}}}}_{\text{intercell interference plus noise}} 
	\end{array}
\end{align}
where ${\mathbf{n}_{{i_k}}} \sim \mathcal{CN}\left\{    \mathbf{0}, \sigma^2\mathbf{I}_{N}  \right\}$ denotes the thermal noise at  user $i_k$. 
From (\ref{received_signal}), the rate of user $ i_k $   can be expressed as \cite{shi2011iteratively}
\begin{align}\label{rate}
	\begin{aligned}
		{R_{{i_k}}} &= \log \det \bigg(   {\mathbf{I}_N} 
		+ {\mathbf{H}_{{i_k},i}}{\mathbf{Q}_i}{\mathbf{T}_i}{\mathbf{W}_{{i_k}}}\mathbf{W}_{{i_k}}^H \mathbf{T}_i^H \mathbf{Q}_i^H \mathbf{H}_{{i_k},i}^H   \\
	&\times	\Big( \sum\nolimits_{\left( {j,u} \right) \ne \left( {i,k} \right)}  \mathbf{J}_{i_k,j,u}  + {\sigma ^2}{\mathbf{I}_N} 
		\Big)^{ - 1}   \bigg),
	\end{aligned}
\end{align}
where $\mathbf{J}_{i_k,j,u}   \triangleq  \mathbf{H}_{{i_k},j}{\mathbf{Q}_j}{\mathbf{T}_j}{\mathbf{W}_{{j_u}}}\mathbf{W}_{{j_u}}^H   \mathbf{T}_j^H  \mathbf{Q}_j^H \mathbf{H}_{{i_k},j}^H $.

We  consider the per-RF chain power constraint rather than the sum-power constraint. As discussed in Section \ref{section0}, this is more realistic as in practical hybrid precoding systems, each RF chain is equipped with its own power amplifier and is limited by the linearity of its amplifier\cite{dsouza2018hybrid}. Based on rate expression (\ref{rate}), the weighted sum user rate maximization problem with per-RF chain  power constraints and metamaterial tunable constraints can be formulated as follows
\begin{subequations}\label{optimization_problem}
	\begin{align} 
		&  \mathop {\max }\limits_{     \{ {\mathbf{W}_{{i_k}}},{\mathbf{Q}_i}\}    }  \;  \;  \;  \sum\nolimits_{i=1}^L {\sum\nolimits_{k = 1}^{{U_i}} {{\omega _{{i_k}}}{    R_{i_k}    }} }   \\
\label{power_constraint}
&\text { s.t. }    \sum\nolimits_{k = 1}^{{U_i}} {   \left[     \mathbf{W}_{i_k}   \mathbf{W}_{i_k} ^H    \right]_{mm} \le {P_{i,m}}},  1\leq m \leq M_y, \forall i, \\\label{HMIMO_constrinat}
&       \qquad   {q_{i,{m_y},{m_x}}} \in {\mathcal{Q}^c},c \in \left\{ {b,g} \right\},\forall i,{m_x},{m_y} ,
	\end{align}
\end{subequations}
where  $  \left[   \mathbf{X}   \right]_{mm}  $ denotes the $m$-th diagonal element of matrix $\mathbf{X}$,  weight $\omega _{{i_k}}$ denotes the priority of  user $i_k$ in the system,  
(\ref{power_constraint}) indicates that the  power amplified by the $ m $-th RF's amplifier of  BS $i$ is not allowed to exceed individual power constraint $P_{i,m}$,
and (\ref{HMIMO_constrinat}) represents the binary tuning constraint ($\mathcal{Q}^b$) or grayscale tuning constraint ($\mathcal{Q}^g$)  of each HMIMO element. 

The problem is non-convex due to the non-convex objective function, non-convex constraints (\ref{HMIMO_constrinat}), and the coupling between  variables, and therefore poses several challenges in designing signal processing algorithms.

\begin{remark}
Design of conventional fully digital and hybrid arrays can be viewed as    special cases of problem (\ref{optimization_problem}). For fully digital array, we have $\mathbf{Q}_i = \mathbf{T}_i = \mathbf{I}_M, \forall i$. For sub-connected hybrid D/A architectures, we can   revise the constraints of  $q_{i,{m_y},{m_x}} $ to be unit-modulus and let $ \mathbf{t}_{i, m_y}^o = \mathbf{1}_{M_x}$, $\forall i, m_y$.
\end{remark}

\section{Multi-Cell Digital Precoding Design}\label{section2}
To begin with, we tackle the non-convex rate function $ {R_{{i_k}}}  $ in  (\ref{rate}) under   the weighted minimum mean square error (WMMSE) and block coordinate descent (BCD) framework\cite{shi2011iteratively}. 
\begin{theorem}
Let ${\mathbf{U}_{{i_k}}}$ and $ \mathbf{V}_{i_k} \succeq \mathbf{0}$ be the linear decoding matrix and the weight matrix for user $i_k$, respectively. Then, problem (\ref{optimization_problem}) is equivalent to \upshape 
\begin{subequations}\label{MSE_problem}
	\begin{align}\label{MSE_objective}
	& \min _{  \{  \mathbf{W}_{i_k}, \mathbf{U}_{i_k}, \mathbf{V}_{i_k}, \mathbf{Q}_i    \}  } 
	\sum_{i=1}^L \sum_{k=1}^{U_i} \omega_{i_k}\left(\operatorname{Tr}\left\{\mathbf{V}_{i_k} \mathbf{E}_{i_k}\right\}-\log \operatorname{det}\left(\mathbf{V}_{i_k}\right)\right) 
	\\
	&\qquad {\text { s.t. } } \qquad ({\rm \ref{power_constraint}}),   ( {\rm \ref{HMIMO_constrinat}}), \nonumber
\end{align}
\end{subequations}
\itshape with the MSE matrix of user $i_k$ given by  
\begin{align}\label{MSE_matrix}
		\mathbf{E}_{i_k}&=\mathbf{I}_d
		-\mathbf{U}_{i_k}^H \mathbf{H}_{i_k, i} \mathbf{Q}_i \mathbf{T}_i \mathbf{W}_{i_k}
		-\left(\mathbf{U}_{i_k}^H \mathbf{H}_{i_k, i} \mathbf{Q}_i \mathbf{T}_i \mathbf{W}_{i_k}\right)^H \nonumber \\
				& +\mathbf{U}_{i_k}^H 
				\Big( \sum\nolimits_{(j, u)} \mathbf{J}_{i_k, j, u}  + {\sigma ^2}{\mathbf{I}_N}    
				 \Big) \mathbf{U}_{i_k},
\end{align}
in the sense that they have the same set of stationary points. \upshape
\end{theorem}

\itshape Proof: \upshape
With linear decoding matrix $\mathbf{U}_{{i_k}}^H$, the estimated signal vector of user $i_k$ can be given by
$ {\hat{ \mathbf{s}}_{{i_k}}} = \mathbf{U}_{{i_k}}^H   {\mathbf{y}_{{i_k}}} $
where ${\mathbf{y}_{{i_k}}}$ is given in (\ref{received_signal}).
Then, the MSE matrix in (\ref{MSE_matrix}) can be obtain by calculating the expectation of  $ {\mathbf{E}_{{i_k}}} \triangleq {\mathbb{E}_{\mathbf{s},\mathbf{n}}}   [ {\left( {{{\hat{ \mathbf{s}}}_{{i_k}}} - {\mathbf{s}_{{i_k}}}} \right){{( \hat {\mathbf{s}}_{{i_k}} - \mathbf{s}_{i_k} )}^H}} ]  $. Given $\mathbf{W}_{i_k}$ and $\mathbf{Q}_i$, problem (\ref{MSE_problem}) is convex with ${\mathbf{U}_{{i_k}}}$ and $ \mathbf{V}_{i_k} $. Thus, letting $\frac{\partial \operatorname{Tr}\left\{\mathbf{V}_{i_k} \mathbf{E}_{i_k}\right\}}{\partial \mathbf{U}_{i_k}}=0$, the optimal variable $ 	\mathbf{U}_{i_k}^{\text {opt}} $ for (\ref{MSE_problem}) is given by \cite[Table 4.3]{hjorungnes2011complex}
\begin{align}\label{optimal_U}
	\begin{aligned}
		\mathbf{U}_{i_k}^{\text {opt}}=\left(\sum\nolimits_{(j, u)} \mathbf{J}_{i_k, j, u}  + {\sigma ^2}{\mathbf{I}_N}    \right)^{-1} \mathbf{H}_{i_k, i} \mathbf{Q}_i \mathbf{T}_i \mathbf{W}_{i_k}.
	\end{aligned}
\end{align}
Also, checking the first-order optimality condition of (\ref{MSE_objective}) with respect to $  \mathbf{V}_{i_k} $ yields 
\begin{align}\label{optimal_V}
\mathbf{V}_{i_k}^{ \text {opt} }=\left(   \mathbf{E}_{i_k}   \right)^{-1}.
\end{align}
Substituting $	\mathbf{U}_{i_k}^{  \text {opt}  } $ and $\mathbf{V}_{i_k}^{  \text {opt}  }$ into problem (\ref{MSE_problem}), it becomes
\begin{subequations}\label{1234}
	\begin{align}\label{1234_objective_funtion}
		& \max _{  \{ \mathbf{W}_{i_k},  \mathbf{Q}_i \}   } 
		\sum\nolimits_{i=1}^L \sum\nolimits_{k=1}^{U_i} \omega_{i_k} \nonumber\\
		&\quad \times 
		\log \operatorname{det}   \left(\left(\mathbf{I}_d-\left(\mathbf{U}_{i_k}^{\text{opt}}\right)^H \mathbf{H}_{i_k, i} \mathbf{Q}_i \mathbf{T}_i \mathbf{W}_{i_k}\right)^{-1}  \right)
		\\
		&\qquad  \text { s.t. }  \qquad ({\rm \ref{power_constraint}}),   ( {\rm \ref{HMIMO_constrinat}}), \nonumber
	\end{align}
\end{subequations}
which is equivalent to problem (\ref{optimization_problem}) after applying the matrix inversion lemma \cite[(2.66)]{hjorungnes2011complex} to (\ref{1234_objective_funtion}).
$ \hfill \blacksquare  $

Then, problem (\ref{optimization_problem}) can be solved by  utilizing the block coordinate descent method to solve problem (\ref{MSE_problem}), which is more tractable. By  fixing three of the variables $\mathbf{U}_{i_k}, \mathbf{V}_{i_k}, \mathbf{W}_{i_k}, \mathbf{Q}_i $,  we will alternately update the fourth. Note that given $\mathbf{W}_{i_k}$ and $\mathbf{Q}_i$, the optimal design of $	\mathbf{U}_{i_k}^{\text {opt}} $ and $\mathbf{V}_{i_k}^{\text {opt}}$ for (\ref{MSE_problem}) have been obtained in (\ref{optimal_U}) and (\ref{optimal_V}). In the following, 
 we  will focus on the optimization with respect to $\mathbf{W}_{i_k}$, given other variables fixed. The optimization with respect to $\mathbf{Q}_i$ will be presented in the next section.

If the sum power constraint is considered, the optimization of baseband precoding in problem (\ref{MSE_problem}) can be  designed with respect to $\mathbf{W}_{i_k}$  by applying the Lagrangian multipliers method in \cite{shi2011iteratively} directly. In the considered  per-RF chain power constraints, we can rewrite  constraint (\ref{power_constraint}) as $\sum_{k=1}^{U_i}\mathrm{Tr}  \left( \mathbf{O}_{m}   \mathbf{W}_{i_k} \mathbf{W}_{i_k}^H\right)    \leq P_{i, m}$, $\forall i, m$, where matrix $ \mathbf{O}_{m} $ is of   all zero elements except $\left[  \mathbf{O}_{m}  \right]_{mm}=1$.  Then, problem (\ref{MSE_problem}) is still a convex quadratically constrained quadratic program problem with respect to $  \mathbf{W}_{i_k}$ and can be solved by interior-point methods. However, due to the $M_y$ quadratic constraints introduced by $\mathbf{O}_m$, the optimal solution of $  \mathbf{W}_{i_k} $  can not be expressed in a simple closed form and a  nested double-loop structure is necessary, which brings high complexity. In the considered multi-cell networks, how to reduce the design complexity of $\mathbf{W}_{i_k}$, $\forall i\in[1,L], k\in[1,U_i]$, under per-RF chain power constraints is of crucial meaning.

Actually, by exploiting the problem structure, closed-form solutions for per-RF precoding variables can be obtained with high computational efficiency in the considered multi-cell scenario. To achieve this goal, we need to split variable $   \mathbf{W}_{i_k}  \in \mathbb{C}^{M_y \times d} $ into $M_y$ vector variables and re-organize the problem. Specifically, define $ \mathbf{W}_{i_k}^H=\left[\mathbf{w}_{i_k, 1}, \mathbf{w}_{i_k, 2}, \ldots, \mathbf{w}_{i_k, M_y}\right] $ corresponding to $M_y$ RF chains.  Stack the precoding of $m$-th RF chain in the $i$-th cell for $U_i$ users as  $ \tilde{\mathbf{w}}_{i, m}=[\mathbf{w}_{i_1, m}^T, \mathbf{w}_{i_2, m}^T, \ldots, \mathbf{w}_{i_{U_i}, m}^T]^T   \in \mathbb{C}^{d U_{i}  \times 1 }$. Then, per-RF chain power constraint (\ref{power_constraint}) can be rewritten as
\begin{align}\label{per_antenna_cons}
\begin{aligned}
\sum\nolimits_{k=1}^{U_i}\left[\mathbf{W}_{i_k} \mathbf{W}_{i_k}^H\right]_{m m} \!\! &=\sum\nolimits_{k=1}^{U_i}\left\|\mathbf{w}_{i_k, m}\right\|^2
=\left\|\tilde{\mathbf{w}}_{i, m}\right\|^2 \\
&\le {P_{i,m}},  \;\;\;\; 1\leq m \leq M_y, \forall i,
\end{aligned}
\end{align}
which provides the constraint with respect to the   design of the $m$-th RF on the $i$-th BS, i.e., $ \tilde{\mathbf{w}}_{i, m} $. With constraint (\ref{per_antenna_cons}), we will apply BCD algorithms to alternately optimize $\mathbf{U}_{i_k}, \mathbf{V}_{i_k}, \mathbf{Q}_i $, and $\tilde{\mathbf{w}}_{i, m}$, $1\leq m \leq M_y$. 

For clarity, let us focus on the design of $m^\prime$-th RF at cell $i^\prime$. Accordingly, we will extract the objective function with respect to variable  $ \tilde{\mathbf{w}}_{i^\prime, m^\prime} $, by fixing $\mathbf{U}_{i_k}, \mathbf{V}_{i_k}, \mathbf{Q}_i $, and $\tilde{\mathbf{w}}_{i, m}$, $\forall i \neq i^\prime, m\neq m^\prime$.  By defining $ \mathbf{T}_i=\left[\mathbf{t}_{i, 1}, \mathbf{t}_{i, 2}, \ldots, \mathbf{t}_{i, M_y}\right] $, we obtain the following result.
\begin{theorem}
Discarding the irrelevant term, the optimization problem with respect to $ \tilde{\mathbf{w}}_{i^\prime, m^\prime}  $ can be formulated as\upshape
\begin{subequations}\label{problem_w_i_m}
	\begin{align}\label{problem_w_i_m_objective_funtion}
		& \min _{\tilde{\mathbf{w}}_{i^\prime, m^{\prime}}} \quad 
\mu_{i^{\prime}, m^{\prime}}\left\|\tilde{\mathbf{w}}_{i^{\prime}, m^{\prime}}\right\|^2+2 \operatorname{Re}\left\{\breve{\mathbf{d}}_{i^{\prime}, m^{\prime}}^H \tilde{\mathbf{w}}_{i^{\prime}, m^{\prime}}\right\}
		\\
		& \text{ s.t. }\left\|\tilde{\mathbf{w}}_{i^{\prime}, m^{\prime}}\right\|^2 \leq P_{i^{\prime}, m^{\prime}},
	\end{align}
\end{subequations}
\itshape where $\mu_{i^{\prime}, m^{\prime}}=\mathbf{t}_{i^{\prime}, m^{\prime}}^H 
\boldsymbol{ \mathcal{ {A}}}_{i^\prime}   \mathbf{t}_{i^{\prime}, m^{\prime}}$, and
\begin{align}\label{miu_ii_mm}
&\boldsymbol{ \mathcal{ {A}}}_{i^\prime}  = \mathbf{Q}_{i^{\prime}}^H  \sum\nolimits_{i=1}^L \sum\nolimits_{k=1}^{U_i}
\omega_{i_k} \mathbf{H}_{i_k, i^\prime}^H \mathbf{U}_{i_k} \mathbf{V}_{i_k} \mathbf{U}_{i_k}^H \mathbf{H}_{i_k, i^{\prime}} 
\mathbf{Q}_{i^\prime}, \nonumber  \\
&\boldsymbol{\mathcal{B}}_{i^\prime} \! =\! 
\operatorname{blkdiag} \! 
\Big(    \omega_{i^\prime_1} \mathbf{H}_{i^\prime_1, i^\prime}^H \mathbf{U}_{i^\prime_1} \mathbf{V}_{i^\prime_1}^H
, ...,  \omega_{i^\prime_{U_{i^\prime}}} \mathbf{H}_{i^\prime_{U_{i^\prime}}, i^\prime}^H \mathbf{U}_{i^\prime_{U_{i^\prime}}} \! \mathbf{V}_{i^\prime_{U_{i^\prime}}}^H \!  \Big) ,  \nonumber\\
&\breve{\mathbf{d}}_{i^\prime, m^{\prime}}^H =\breve{\mathbf{c}}_{i^\prime, m^{\prime}}^H 
-\left(\mathbf{1}_{U_{i^\prime}}^T \otimes \mathbf{t}_{i^\prime, m^\prime}^H \mathbf{Q}_{i^{\prime}}^H\right) \boldsymbol{\mathcal{B}}_{i^\prime },
\nonumber \\
&\breve{\mathbf{c}}_{i^{\prime}, m^{\prime}}^H =\mathbf{t}_{i^{\prime}, m^{\prime}}^H 
\boldsymbol{ \mathcal{ {A}}}_{i^\prime}  
\sum\nolimits_{{m=1, m \neq m^{\prime}}}^{M_y}  \mathbf{t}_{i^\prime, m} \tilde{\mathbf{w}}_{i^\prime, m}^H.
\end{align}
\end{theorem}\upshape

\itshape Proof: \upshape   The optimization problem with respect to $ \tilde{\mathbf{w}}_{i^\prime, m^\prime}  $ can be formulated as
\begin{subequations}\label{problem_VE}
	\begin{align}\label{problem_VE_objective_funtion}
			& \min _{\tilde{\mathbf{w}}_{i^\prime, m^{\prime}}} \quad \sum\nolimits_{i=1}^L \sum\nolimits_{k=1}^{U_i} \omega_{i_k} \operatorname{Tr}\left\{\mathbf{V}_{i_k} \mathbf{E}_{i_k}\right\} \\
			& \text { s.t. }\left\|\tilde{\mathbf{w}}_{i^{\prime}, m^{\prime}}\right\|^2 \leq P_{i^{\prime}, m^{\prime}},
	\end{align}
\end{subequations}
where $ \mathbf{E}_{i_k} $ is given in (\ref{MSE_matrix}). Then, discarding the constant term, the objective function (\ref{problem_VE_objective_funtion}) can be expanded as
\begin{align}\label{w_i_m_objective_func}
	& \max _{\tilde{\mathbf{w}}_{i^\prime, m^{\prime}}} 
		\sum_{i=1}^L \sum_{k=1}^{U_i} 
		\! -2 \omega_{i_k} \!\!\operatorname{Tr}\!\!
		\left(    \!\!   \operatorname{Re} \!\!
		\left\{ \!\!
		\mathbf{V}_{i_k} \mathbf{U}_{i_k}^H \mathbf{H}_{i_k, i} \mathbf{Q}_i   \!\!
		\sum_{m=1}^{M_y} 
		\! \mathbf{t}_{i, m} \mathbf{w}_{i_k, m}^H
\!	\!	\right\}
	\!	\right)  \nonumber \\
		& +\sum_{i=1}^L \sum_{k=1}^{U_i} \omega_{i_k} \operatorname{Tr}\bigg(\mathbf{V}_{i_k} 
		\mathbf{U}_{i_k}^H 
		\sum_{(j, u)} \mathbf{H}_{i_k, j} \mathbf{Q}_j \sum_{m_1=1}^{M_y} \mathbf{t}_{j, m_1} \mathbf{w}_{j_u, m_1}^H
	  \nonumber   	\\
		 & \times \sum\nolimits_{m_2=1}^{M_y} \mathbf{w}_{j_u, m_2} \mathbf{t}_{j, m_2}^H \mathbf{Q}_j^H \mathbf{H}_{i_{k, j}}^H \mathbf{U}_{i_k}
		 \bigg),
\end{align}
in which the property of $\mathbf{T}_i \mathbf{W}_{i_k}=\sum\nolimits_{m=1}^{M_y} \mathbf{t}_{i, m} \mathbf{w}_{i_k, m}^H$ is used.
The proof then follows by extracting and re-organizing the terms relevant to  $ \tilde{\mathbf{w}}_{i^\prime, m^\prime}  $ from (\ref{w_i_m_objective_func}). For example, relevant terms from the first row of (\ref{w_i_m_objective_func}) can be processed as
\begin{align}
	& -\sum_{m=m^{\prime}}^{M_y} \sum_{i=i^{\prime}}^L \sum_{k=1}^{U_{i^\prime}} 2 \omega_{i_k} \operatorname{Re}\left\{
	\mathbf{t}_{i, m}^H \mathbf{Q}_i^H \mathbf{H}_{i_k, i}^H \mathbf{U}_{i_k} \mathbf{V}_{i_k}^H \mathbf{w}_{i_k, m}
	\right\} \nonumber \\
	& =-\sum\nolimits_{k=1}^{U_{i^\prime}} 2 \omega_{i^\prime_k} \operatorname{Re}\left\{\mathbf{t}_{{i^\prime}, m^{\prime}}^H \mathbf{Q}_{i^{\prime}}^H \mathbf{H}_{i^{\prime}_k, i^{\prime}}^H \mathbf{U}_{i^{\prime}_k} \mathbf{V}_{i^{\prime}_k}^H \mathbf{w}_{i^{\prime}_k, m^{\prime}}\right\}
	\nonumber \\
	& = -2 \operatorname{Re}\left\{\left(\mathbf{1}_{U_{i^{\prime}}}^T \otimes \mathbf{t}_{i^{\prime}, m^{\prime}}^H \mathbf{Q}_{i^{\prime}}^H\right) \boldsymbol{\mathcal{B}}_{i^{\prime}} \tilde{\mathbf{w}}_{i^{\prime}, m^{\prime}}\right\}.
\end{align}
After some algebraic manipulation for the next terms of (\ref{w_i_m_objective_func}), we can complete the proof.
$ \hfill \blacksquare $

 Problem (\ref{problem_w_i_m}) is tractable and admits a closed-form solution. Since $\mu_{i^{\prime}, m^{\prime}}\ge 0$, for the non-trivial case of $\mu_{i^{\prime}, m^{\prime}}>0$, (\ref{problem_w_i_m}) is equivalent to 
 \begin{align}
\min _{\tilde{\mathbf{w}}_{i^\prime, m^{\prime}}}\left\|
\tilde{\mathbf{w}}_{i^{\prime}, m^{\prime}}+\frac{\breve{\mathbf{d}}_{i^{\prime}, m^{\prime}}}{{\mu_{i^{\prime}, m^{\prime}}}}\right\|^2  
\; \text {s.t. }\left\|\tilde{\mathbf{w}}_{i^{\prime}, m^{\prime}}\right\|^2 \leq P_{i^{\prime}, m^{\prime}}.
 \end{align}
Based on triangle inequality, we have $ |\| \mathbf{x}\|-\| \mathbf{y}\|| \leq\|\mathbf{x}+ \mathbf{y}\| $ and the equality holds when $\mathbf{x}= - \varsigma  \mathbf{y}$, $\varsigma\geq 0$. Thus, we obtain the following optimal solution for problem (\ref{problem_w_i_m}):
\begin{align}\label{optimal_w}
\tilde{\mathbf{w}}_{i^{\prime}, m^{\prime}}^{\text {opt }}
=- \min \left\{\frac{1}{\mu_{i^{\prime}, m^{\prime}}}, \frac{\sqrt{P_{i^{\prime}, m^{\prime}}}}
{\left\|\breve{\mathbf{d}}_{ i^{\prime}, m^{\prime}   }\right\|_2}\right\} \breve{\mathbf{d}}_{i^{\prime}, m^{\prime}},
\end{align}
which can be computed efficiently without matrix inversion and bisection search. The computational complexity for obtaining $\tilde{\mathbf{w}}_{i^{\prime}, m^{\prime}}^{\text {opt }}$ in (\ref{optimal_w}) mainly comes from calculating parameters  $ \breve{\mathbf{c}}_{i^{\prime}, m^{\prime}}^H  $ and $\boldsymbol{ \mathcal{ {A}}}_{i^\prime} $ in (\ref{miu_ii_mm}), which is on the order of $ \mathcal{O} (M^2 + M  d \overline{U} )$, by assuming $U_i = \overline{U}, \forall i$. Thus, the overall complexity for obtaining all digital precoder, i.e., $\tilde{\mathbf{w}}_{i^{\prime}, m^{\prime}}^{\text {opt }}$, $\forall i^\prime, m^\prime$, is $\mathcal{O} (L  M^2  + L M  d \overline{U})$.  By contrast, the interior-point method-based algorithm for solving problem (\ref{MSE_problem}) with respect to $   \mathbf{W}_{i_k} $,  $\forall i_k$, needs a much higher complexity of $\mathcal{O} ( L M^2   \overline{U}  +  (L  M_y d   \overline{U}  )^{3.5}   )$ due to the number of variables in $   \mathbf{W}_{i_k} $,  $\forall i_k$, and the calculations in $   \mathbf{E}_{i_k} $,  $\forall i_k$.

\section{HMIMO Antenna Response Design}\label{section3}
In this section, we design the HMIMO tunable response matrix $ \mathbf{Q}_i $ with  other variables fixed. To begin with, we need to rewrite the objective function favorable for exploiting the diagonal feature of $ \mathbf{Q}_i $, as follows
\begin{align}\label{Tr_VE}
\begin{aligned}
&\operatorname{Tr}\left\{\mathbf{V}_{i_k} \mathbf{E}_{i_k}\right\}=\operatorname{Tr}
\bigg\{- 2 \operatorname{Re}\left\{ {\mathbf{T}}_i \mathbf{W}_{i_k} \mathbf{V}_{i_k} \mathbf{U}_{i_k}^H \mathbf{H}_{i_k, i} {\mathbf{Q}}_i \right\}\\
&+\sum\limits_{(j, u)} \mathbf{H}_{i_k, j}^H \mathbf{U}_{i_k} \mathbf{V}_{i_k} \mathbf{U}_{i_k}^H \mathbf{H}_{i_k, j} {\mathbf{Q}}_j {\mathbf{T}}_j \mathbf{W}_{j_u} \mathbf{W}_{j_u}^H {\mathbf{T}}_j^H {\mathbf{Q}}_j^H
\bigg\}+c\\
&=  
\sum_{(j,u)} 
\operatorname{Tr}\left\{\mathbf{F}_{i_k, j} {\mathbf{Q}}_j \mathbf{F}_{j, u} {\mathbf{Q}}_j^H \right\}
- 2 \operatorname{Re}
\left\{
\operatorname{Tr}\left\{\mathbf{F}_{i_k} {\mathbf{Q}}_i\right\}
\right\}
 +c 
\end{aligned}
\end{align}
where
\begin{align}\label{Fs}
	\begin{aligned}
		& \mathbf{F}_{i_k}={\mathbf{T}}_i \mathbf{W}_{i_k} \mathbf{V}_{i_k} \mathbf{U}_{i_k}^H \mathbf{H}_{i_k, i} ,\\
		& \mathbf{F}_{i_k, j}=\mathbf{H}_{i_k, j}^H \mathbf{U}_{i_k} \mathbf{V}_{i_k} \mathbf{U}_{i_k}^H \mathbf{H}_{i_k, j} \succeq \mathbf{0},\\
		& \mathbf{F}_{j, u}={\mathbf{T}}_j \mathbf{W}_{j_u} \mathbf{W}_{j_{u}}^H {\mathbf{T}}_j^H \succeq \mathbf{0}.
	\end{aligned}
\end{align}

Recall  that $ {\mathbf{Q}}_i=\operatorname{diag}\left\{\boldsymbol{ \mathbf{q}}_i\right\} $. Using  properties of the diagonal matrix\cite[(165)]{zhiTwotimescale2022}, we have
$ \operatorname{Tr}\left\{\mathbf{A} {\mathbf{Q}}_i \mathbf{B} {\mathbf{Q}}_i^H\right\}=\boldsymbol{\mathbf{q}}_i^H\left(\mathbf{A} \odot \mathbf{B}^T\right) \boldsymbol{\mathbf{q}}_i $ and 
$  \operatorname{Tr}\{\mathbf{A}{\mathbf{Q}}_i \}=\operatorname{diag}\{\mathbf{A}\}^T \boldsymbol{\mathbf{q}}_i$. Then, we can process (\ref{Tr_VE}) to a quadratic form:
\begin{align}\label{rewrite_VE}
\begin{aligned}
	 \operatorname{Tr}\left\{\mathbf{V}_{i_k} \mathbf{E}_{i_k}\right\}
	 &=\sum\nolimits_{j=1}^{L} \boldsymbol{\mathbf{q}}_j^H\left(\sum\nolimits_{u=1}^{U_j} \mathbf{F}_{i_k, j} \odot \mathbf{F}_{j, u}^T\right) \boldsymbol{\mathbf{q}}_j\\
	 &-2 \operatorname{Re}\left\{\operatorname{diag}\left\{\mathbf{F}_{i_k}\right\}^T \boldsymbol{\mathbf{q}}_i\right\}+\mathfrak{c}.
\end{aligned}
\end{align}
Then, the optimization of HMIMO tunable response can be decoupled between different cells. For the sake of clarity, let us focus on  the  cell $i'$. Substituting (\ref{rewrite_VE}) into objective function (\ref{MSE_objective}),  optimization problem (\ref{MSE_problem})  with respect to $ \mathbf{q}_{i^{\prime}}  $ can be formulated compactly as
\begin{subequations}\label{optimization_HMIMO_phase}
	\begin{align}\label{30a}
	&\min\limits_{    \mathbf{q}_{i^{\prime}}    }  f_o\left(\mathbf{q}_{i^{\prime}}\right)=\mathbf{q}_{i^{\prime}}{ }^H \mathbf{A}_o \mathbf{q}_{i^{\prime}}-2 \operatorname{Re}\left\{\mathbf{b}_o^T \mathbf{q}_{i^{\prime}}\right\}
	+ \hat{ \mathfrak{c}}
	   \\\label{constraint_HMIMO_phase}
&\text { s.t. } \quad\left[\boldsymbol{   \mathbf{q}}_{i^\prime}   \right]_m \in \mathcal{Q}^c, c \in\{b, g\}, 1 \leq m \leq M,
	\end{align}
\end{subequations}
where
\begin{align}
\mathbf{b}_o^T&=\sum\nolimits_{k=1}^{U_{i^\prime }} \omega_{i^\prime_k} \operatorname{diag}\big\{\mathbf{F}_{i_k^{\prime}}\big\}^T,
\\\label{Aos}
\mathbf{A}_o&=\sum\nolimits_{i=1}^L \sum\nolimits_{k=1}^{U_i} \omega_{i_k}\big(
\sum\nolimits_{u=1}^{U_{i'}} \mathbf{F}_{i_k, i^{\prime}} \odot \mathbf{F}_{i^{\prime}, u}^T
\big).
\end{align}
It can be seen that the reformulated problem has a convex objective function with respect to $\boldsymbol{\mathbf{q}}_{i'}$. Therefore, we only need to tackle the non-convex constraints. As baseline algorithms, 
the most simple way is to first solve the unconstrained problem and then project the obtained solution to the closest point in $\mathcal{Q}^b$ or $\mathcal{Q}^g$. In this case, we can derive the first-order derivative of the objective function in (\ref{optimization_HMIMO_phase}) as 
$  {\partial f_o\left(\mathbf{q}_{i^{\prime}}\right)}/{\partial \mathbf{q}_{i^{\prime}}}=\mathbf{A}_o^T \mathbf{q}_{i^{\prime}}^*-\mathbf{b}_o$ \cite[Table 4.3]{hjorungnes2011complex}.
Letting it equal to zero yields the unconstrained optimal solution $  {\boldsymbol{\mathbf{q}}}_{i^\prime}^{\text{opt}}=\left(\mathbf{A}_o^H\right)^{-1} \mathbf{b}_o^* $. To satisfy the binary constraints, each element in $  {\boldsymbol{\mathbf{q}}}_{i^\prime}^{\text{opt}} $ can be projected to the nearest point in $\mathcal{Q}^b$, leading to
\begin{align}\label{simple_binary}
 \tilde{\mathbf{Q}}_{i^\prime}^{\text {b}}=\operatorname{diag}\left\{f_{+}\left\{\operatorname{Re}\left\{
 {\boldsymbol{\mathbf{q}}}_{i^\prime}^{\text{opt}}
\right\}- \varpi\mathbf{1}/2    \right\}\right\},
\end{align}
where $f_{+}\{x\}=0 $ if $ x<0 $ and  $  f_{+}\{x\}=\varpi  $   if  $ x \geq 0 $. Under Lorentzian-constrained phase model, each element in $  {\boldsymbol{\mathbf{q}}}_{i^\prime}^{\text{opt}}   $ is projected to the nearest point in $\mathcal{Q}^g$, resulting in 
\begin{align}\label{simple_gray}
\tilde{\mathbf{Q}}_{i^\prime}^{ {g }}=\frac{1}{2}\operatorname{diag}\left\{\exp\left\{j \angle\left(
 {\boldsymbol{\mathbf{q}}}_{i^\prime}^{\text{opt}}
-   j\mathbf{1}/2    \right)\right\}+j \mathbf{1}\right\}.
\end{align}
While simplest, the solutions in (\ref{simple_binary}) and (\ref{simple_gray}) cannot guarantee  performance and may generate severe performance loss. In the following, we will propose  more sophisticated algorithms.

\subsection{Design with Binary Constraints}
In this section, we will tackle the optimization problem with the binary constraint, i.e., with constraint $ \mathcal{Q}^b$ in (\ref{constraint_HMIMO_phase}). This discrete constraint makes the finding of optimal solutions  highly challenging. 

For tractability, we first present the following lemma.
\begin{lem}
Define $\breve{\mathbf{q}}_{i^{\prime}}=(2 \mathbf{q}_{i^{\prime}}-\varpi \mathbf{1})/\varpi  $. Problem (\ref{optimization_HMIMO_phase}) is equivalent to the following real optimization problem:
	\begin{align} \label{quadratic_form}
	\min\limits_{   \breve { \mathbf{q}}_{i^{\prime}}   \in \{-1, 1\}^M  }  F_r\left(   \breve{\mathbf{q}}_{i^{\prime} }    \right)  \triangleq  \breve{\mathbf{q}}_{i^{\prime}}{ }^T \mathbf{A} \breve{\mathbf{q}}_{i^{\prime}}-2 \mathbf{b}^T \breve{\mathbf{q}}_{i^{\prime}} ,
\end{align}
where $	\mathbf{A}   =  \varpi^2 \operatorname{Re}\left\{\mathbf{A}_o\right\} /4  $ and $ \mathbf{b}^T ={\varpi}{} \operatorname{Re}\left\{\mathbf{b}_o^T\right\}/2- {\varpi^2}{} \mathbf{1}^T   \operatorname{Re}\left\{\mathbf{A}_o\right\}/4$.
\end{lem}

\itshape Proof: \upshape 
Since $ \mathbf{F}_{i_k, j}, \mathbf{F}_{j, u} \succeq \mathbf{0}$, we have $\mathbf{A}_o \succeq \mathbf{0}$ and $ \hat{ \mathfrak{c}}    \in \mathbb{R}$. Thus,   objective function$  f_o\left(\mathbf{q}_{i^{\prime}}\right)  $ in (\ref{30a}) yields real-value results so that $   f_o\left(\mathbf{q}_{i^{\prime}}\right)  = \mathrm{Re}\left\{   f_o\left(\mathbf{q}_{i^{\prime}}\right)  \right\}  $.  As a result, we can define $ F_r\left(\mathbf{q}_{i^{\prime}}\right) \triangleq    \mathrm{Re}\left\{     f_o\left(\mathbf{q}_{i^{\prime}}\right)  \right\}$ as the new objective function.  Substituting the definition of $ \mathbf{q}_{i^{\prime}}=(\varpi \breve{\mathbf{q}}_{i^{\prime}}+\varpi \mathbf{1})/2$ into real function $ F_r\left(\mathbf{q}_{i^{\prime}}\right)  $ and removing the constant terms, we can complete the proof after some simplifications with the new variable of $\left[\breve{\mathbf{q}}_{i^{\prime}}\right]_m \in \{-1, 1\}, \forall  m $, by using the properties that ${\mathbf{q}}_{i^\prime}$ is  real, $ \operatorname{Re}\left\{\mathbf{q}_{i^{\prime}}^T \mathbf{A}_o \mathbf{q}_{i^{\prime}}\right\}=\mathbf{q}_{i^{\prime}}^T\operatorname{Re}\left\{\mathbf{A}_o\right\} \mathbf{q}_{i^{\prime}} $, and $ \left\{\operatorname{Re}\left\{\mathbf{A}_o\right\}\right\}^T=\left\{\operatorname{Re}\left\{\mathbf{A}_o\right\}\right\}^H=\left\{\operatorname{Re}\left\{\mathbf{A}_o\right\}\right\} $.
$ \hfill \blacksquare $

Problem (\ref{quadratic_form}) is NP-hard and its solution based on exhaustive search yields a complexity of $\mathcal{O}(2^M)$. How to reduce the computational complexity is of practical meaning. In the following, we will provide two effective algorithms for solving (\ref{quadratic_form}), respectively.

\subsubsection{Hidden Convexity-Based Algorithm}
We first propose a low-complexity sub-optimal algorithm for   problem (\ref{quadratic_form}) which yields closed-form solutions. Specifically, compared with relaxing $ \breve{\mathbf{q}}_{i^{\prime}   }$ to the whole space $\mathbb{R}^M$ (i.e., without constraints), we  consider a tighter relaxation of constant-norm constraint $\| \breve{\mathbf{q}}_{i^{\prime}}\|^2=M$ and relax the optimization problem (\ref{quadratic_form}) as
\begin{align}\label{problem_relax_quadratic}
\min _{\left\|   \breve{\mathbf{q}}_{i^\prime}\right\|^2=M}   
\breve{\mathbf{q}}_{i^{\prime}}{ }^T \mathbf{A} \breve{\mathbf{q}}_{i^{\prime}}
-2 \mathbf{b}^T\breve{\mathbf{q}}_{i^{\prime}}.
\end{align}
It can be seen that the considered relaxation is theoretically tighter than the common  relaxation of $\left\|  \breve{\mathbf{q}}_{i^\prime}\right\|^2\leq M$. However, it  makes   problem (\ref{problem_relax_quadratic}) non-convex due to the non-convex constant-norm constraint. In the following, we will tackle this challenge by leveraging the hidden convexity property \cite{ben1996hidden,eldar2005hidden}  in the quadratic problem.

Note that $\mathbf{A} \succeq \mathbf{0}$ is a real symmetric matrix. Therefore, applying the eigenvalue decomposition, we have $\mathbf{A}  = \mathbf{S}\mathbf{\Lambda }\mathbf{S}^T$, where $\mathbf{\Lambda }=\mathrm{diag}\left\{\lambda_1,\cdots,\lambda_M\right\}$ is a diagonal matrix comprised of the eigenvalues   $\lambda_m \geq 0$, $\forall m$. Now, let $\hat{\mathbf{q}}_{i^{\prime}} = \mathbf{S}^T \breve{\mathbf{q}}_{i^{\prime}}   $. Due to the property $\mathbf{S}^{-1} = \mathbf{S}^T$ and $\left\|\hat{\mathbf{q}}_{i^{\prime}}\right\|^2=\left\|  \breve{\mathbf{q}}_{i^{\prime}}\right\|^2=M$, the problem in (\ref{problem_relax_quadratic}) can be rewritten with respect to the new variable $ \hat{\mathbf{q}}_{i^{\prime}} $ as follows
\begin{align}\label{problem_EVD}
\min _{\left\|\hat{\mathbf{q}}_{i^{\prime}}\right\|^2=M} \hat{\mathbf{q}}_{i^{\prime}}^T \mathbf{\Lambda} \hat{\mathbf{q}}_{i^{\prime}}-2 \hat{\mathbf{b}}^T \hat{\mathbf{q}}_{i^{\prime}} 
=\sum\nolimits_{i^{\prime}=1}^M \lambda_{i^{\prime}} \hat{q}_{i^{\prime}}^2-2 b_{i^{\prime}} \hat{q}_{i^{\prime}},
\end{align}
where  $\hat{q}_{i^{\prime}} = [ \hat{\mathbf{q}}_{i^{\prime}} ]_{i^{\prime}} =[ \mathbf{S}^T \breve{\mathbf{q}}_{i^{\prime}} ]_{i^{\prime}}$, $ \hat{\mathbf{b}}^T=\mathbf{b}^T \mathbf{S} $, and ${b}_{i^{\prime}} = [ \hat{\mathbf{b}} ]_{i^{\prime}} $.

To transfer  problem (\ref{problem_EVD}) to a convex form, we need the following property of the quadratic problem\cite{ben1996hidden}.
\begin{lem}\label{lemma_necessary}
	If $\mathbf{x}^{\star}=[x^{\star}_1,\cdots, x^{\star}_M]^T$ is a globally optimal solution of quadratic problem $   \mathbf{x} ^T \mathbf{\Lambda}\mathbf{x}-2 \hat{\mathbf{b}}^T \mathbf{x}$ subject to $\|\mathbf{x}\|^2=M$, it must satisfy the condition of
	 $ x^{\star}_{i^{\prime}} b_{i^{\prime}} \geq 0, \forall i^{\prime}\in[1,M]$.
\end{lem}

Leveraging Lemma \ref{lemma_necessary},  by defining variable $\hat{q}_{i^\prime}=\operatorname{sign}\left(b_{i^{\prime}}\right) \sqrt{\theta_{i^\prime}}$, we can  address the non-convex  constant-norm constraint $ \left\|\hat{\mathbf{q}}_{i^{\prime}}\right\|^2=M $ and recast problem  (\ref{problem_EVD}) as follows
\begin{subequations}\label{convex_problem}
	\begin{align} 
	& \min _{\theta_{i^\prime}} \sum\nolimits_{i^{\prime}=1}^M \lambda_{i^{\prime}} \theta_{i^{\prime}}-2\left|b_{i^{\prime}}\right| \sqrt{\theta_{i^{\prime}}} \\
& \text { s.t. } \theta_{i^{\prime}} \geq 0, \forall i^{\prime}, \\\label{sum_constraint}
&\quad  \sum\nolimits_{i^{\prime}=1}^M \theta_{i^{\prime}}=M,
	\end{align}
\end{subequations}
where $ \operatorname{sign}\left(b_{i^{\prime}}\right)=1$, if $b_{i^{\prime}} \geq 0 $, and $ \operatorname{sign}\left(b_{i^{\prime}}\right)=-1$, if $b_{i^{\prime}} < 0 $. After exploiting the optimal condition, it can be observed that problem ($\ref{convex_problem}$) is now convex. It can be solved in closed form by using the Lagrange multiplier method. Specifically, attaching a Lagrange multiplier $\mu$ to constraint (\ref{sum_constraint}), we got the following Lagrange function
\begin{align}\label{Lagrange_function}
L\!\!\left( {\left\{ {{\theta _{i'}}} \right\}_{i' = 1}^M,\mu }\! \right) \!=\! \sum\limits_{i' = 1}^M {{\lambda _{i'}}{\theta _{i'}}}  - 2\left| {{b_{i'}}} \right|\sqrt {{\theta _{i'}}}  + \mu \left( {\sum\limits_{i' = 1}^M {{\theta _{i'}} - } M}\!\! \right).
\end{align}
Letting $ \frac{{\partial L\left( {\left\{ {{\theta _{i'}}} \right\}_{i' = 1}^M,\mu } \right)}}{{\partial {\theta _{i'}}}} =0 $, we have $ \sqrt {{\theta _{i'}}}  = \frac{{\left| {{b_{i'}}} \right|}} {{{\lambda _{i'}} + \mu }}\ge 0$. Therefore, we obtain
\begin{align}\label{optimal_theta}
	& \theta_{i^{\prime}}^{\text{opt}}=\frac{b_{i^{\prime}}^2}{\left(\lambda_{i^{\prime}}+\mu\right)^2}\ge 0, \forall i^\prime , \\
	& \mu\ge -\lambda_{i^{\prime}}, \forall i^\prime.
\end{align}

Substituting (\ref{optimal_theta}) into Lagrange function (\ref{Lagrange_function}), the dual function is given by
\begin{align}
g\left( \mu  \right) = \mathop {\min }\limits_{{\theta _{i'}} \geqslant 0} L\left( {\left\{ {{\theta _{i'}}} \right\}_{i' = 1}^M,\mu } \right) =  - \sum\nolimits_{i' = 1}^M {\frac{{b_{i'}^2}}{{ {{\lambda _{i'}} + \mu } }}}  - \mu M.
\end{align}
The dual problem of (\ref{convex_problem}) is 
\begin{align}\label{dual_function_mu}
\mathop {\max }\limits_\mu  \left\{ {g\left( \mu  \right):\mu  \geqslant {-\lambda _{i'}},\forall i'} \right\}.
\end{align}
 $ g\left( \mu  \right) $ is a convex function with respect to $\mu$. By letting the first-order derivative be zero, we have condition $ \sum\nolimits_{i^{\prime}=1}^M b_{i^{\prime}}^2 /\left(\lambda_{i^{\prime}}+\mu\right)^2=M $ for the optimal value of $\mu$, which  can be 
easily found through bisection search between $(\max\left\{-\lambda_{i^{\prime}}\right\}, +\infty)$. After find the optimal value of $\mu^{\text{opt}}$, we have $\theta_{i^{\prime}}^{\text{opt}}=\frac{b_{i^{\prime}}^2}{\left(\lambda_{i^{\prime}}+\mu^{\text{opt}}\right)^2}$ and $\hat{q}_{i^\prime}^{\text{opt}}=\operatorname{sign}\left(b_{i^{\prime}}\right) \sqrt{\theta_{i^\prime}^{\text{opt}}}$. By constructing vector $\hat{ \mathbf{q}}_{i'}^{\text{opt}} =[\hat{q}_{1}^{\text{opt}},\cdots,\hat{q}_{M}^{\text{opt}}] ^T $,  the  solution of  problem (\ref{quadratic_form}) can be given by
\begin{align}\label{optimal_Q_hidden}
\breve{\mathbf{q}}_{i'}^{\text{opt}} \! = \!\!  \begin{cases}     
	\mathcal{P}_b\left\{\mathbf{S}    \hat {\mathbf{q}}_{i'}^{\text{opt}}       \right\}, & \text { if }  F_r\left(   	\mathcal{P}_b\left\{\mathbf{S}\hat {\mathbf{q}}_{i'}^{\text{opt}}\right\}      \right)    \le   F_r\left(  [\breve{\mathbf{q}}_{i'}]^{\text{former}}     \right) ,  \\ 
	[\breve{\mathbf{q}}_{i'}^{\text{opt}}]^{\text{former}}, & \text { otherwise },\end{cases}
\end{align}
where $ \mathcal{P}_b $ is a projection operator onto closest points in binary sets $\left\{-1, +1\right\}^M$, $ \mathbf{S}\hat {\mathbf{q}}_{i'}^{\text{opt}} $ obtains the optimal solution of problem (\ref{problem_relax_quadratic}) since $\mathbf{S} \hat{\mathbf{q}}_{i^{\prime}} =\mathbf{S}  \mathbf{S}^T \breve{\mathbf{q}}_{i^{\prime}}  =  \breve{\mathbf{q}}_{i^{\prime}}   $,
$F_r(\cdot)$ is the objective function of (\ref{quadratic_form}), and $	[\breve{\mathbf{q}}_{i'}]^{\text{former}}$ denotes the optimized result  of $ \breve{\mathbf{q}}_{i'} $ obtained in the former iteration. With  (\ref{optimal_Q_hidden}), we can guarantee the non-increasing feature of the objective function without the influence of the projection operation.
Next, the local search method \cite{papadimitriou1998combinatorial} with respect to each individual element in  $  \breve{\mathbf{q}}_{i'}^{\text{opt}}    $ can be used to further improve the performance with  a complexity of $\mathcal{O}(M)$.
Finally, the solution of problem (\ref{optimization_HMIMO_phase})  under constraint $\left\{0, \varpi\right\}^M$ can be obtained    by  $\left[\boldsymbol{\mathbf{q}}_{i^{\prime}}^{\text{opt}}    \right]_m  ={  ( \varpi\left[\breve{\mathbf{q}}_{i^{\prime}}^{\text{opt}}        \right]_m +\varpi)/2 }$, $\forall m$.

Due to the eigenvalue decomposition of matrix $\mathbf{A}$, the hidden-convexity-based method requires a complexity of $\mathcal{O}(LM^3)$ for tackling problem (\ref{quadratic_form}) of all $L$ cells. Besides, computing   all parameters   in (\ref{Fs}) and (\ref{Aos})
mainly result in complexity of $\mathcal{O}( L^2 \overline{U}  M^2 d  +  L \overline{U}^2 M^2       ) $.

\subsubsection{Sphere Decoding-Based Algorithms} 
The proposed HC algorithm yields sub-optimal solutions for problem (\ref{optimization_HMIMO_phase}) with low complexity. To further improve the performance and also enable us to evaluate the effectiveness of the proposed HC algorithm, in this subsection, without any relaxation, we would like to obtain  the global-optimal solution for problem (\ref{optimization_HMIMO_phase}), by proposing an accelerated SD algorithm.

Specifically, applying the Cholesky decomposition to matrix $\mathbf{A}^{\prime} \triangleq  \mathbf{A}+\zeta  \mathbf{I}_M \succ \mathbf{0}$, we have $\mathbf{A}^{\prime}    =\mathbf{C}^T  \mathbf{C}   $, where $\mathbf{C}$ is a real upper triangular matrix. Since $\left\|   \breve{\mathbf{q}}_{i^\prime}\right\|^2=M$, problem (\ref{quadratic_form}) is equivalent to
\begin{align}\label{d}
		\begin{aligned}
&\mathop {\min }\limits_{{  \breve{\mathbf{q}}_{i'}} =    {{\left\{ {    -    1, + 1    } \right\}}   ^M}}    \breve{\mathbf{q}}_{i^{\prime}}{ }^T \mathbf{A} \breve{\mathbf{q}}_{i^{\prime}}   -    2 \mathbf{b}^T \breve{\mathbf{q}}_{i^{\prime}}    +    \zeta M \\	
&\quad\quad\;\;  =
\breve{\mathbf{q}}_{i^{\prime}}{ }^T \mathbf{A}^{\prime} \breve{\mathbf{q}}_{i^{\prime}}-2\mathbf{b}^T \breve{\mathbf{q}}_{i^{\prime}} =\left\|\mathbf{C}\breve{\mathbf{q}}_{i^{\prime}}-\mathbf{d}\right\|^2-\|\mathbf{d}\|^2    
		\end{aligned}
\end{align}
where
$ 	\mathbf{d} \triangleq \left(\mathbf{C}^{-1}\right)^{T} \mathbf{b} $.
For expression clarity,  we omit the subscript $i^\prime$ of $\mathbf{q}_{i^{\prime}}$ in the sequel of this subsection and denote  problem (\ref{d})  as
	\begin{align} \label{SD_optimization_problem}
		\mathop {\min }\limits_{{\mathbf{q} } = {{\left\{ { - 1, + 1} \right\}}^M}}   \left\|\mathbf{C} \mathbf{q}-\mathbf{d}\right\|^2,
	\end{align}
whose globally optimal solution can be obtained by exploiting SD methods\cite{giuseSD2003,RamezaniSD2024,stojnic2008speeding,hassibi2005sphere}.  
 SD investigates all the points $\mathbf{q} = {{\left\{ { - 1, + 1} \right\}}^M}$ that make lattice points $\mathbf{C} \mathbf{q}$ lie in an $M$-dimensional hypersphere whose center is $\mathbf{d}$ and radius is $r$. Specifically, SD search over all points ${\mathbf{q}} $ satisfying
\begin{align}\label{sphere_condition}
\left\|\mathbf{C} \mathbf{q}- \mathbf{d}\right\|^2 \leq r^2
\end{align}
and then selects the one with minimal objective values. Clearly, the closest lattice point inside the sphere will also be the closest lattice point for the whole lattice.

The real-value radius $r$ should be properly selected. If $r$ is too large, all the lattice points will be investigated and the SD will degrade to the brute force search. If $r$ is too small, no point will be found within the sphere. The computational efficiency of SD  depends on the choice of $r$ and we hope the number of candidate points within the sphere to be sufficiently small.  Depending on the proper selection of an efficient radius, SD can avoid the exhaustive search and achieve the optimal solution with much-reduced search space and computational complexities\footnote{We conduct the HC algorithm to determine the initial value of $r$, taking advantage of the algorithm's simplicity and effectiveness.}.

SD will be carried out in a branch-and-bound way.  Due to the upper triangular feature of $\mathbf{C}$,   it can be divided into four blocks and denoted  by 
\begin{align}
	\mathbf{C}=\left[\begin{array}{cc}
		\mathbf{C}_{1: k-1,1: k-1} & \mathbf{C}_{1: k-1, k: M} \\
		\mathbf{0}^T & \mathbf{C}_{k: M, k: M}
	\end{array}\right]
\end{align}
where $\left[  \mathbf{C}\right]_{x_1:y_1, x_2:y_2}  \in \mathbb{R}^{(y_1-x_1+1)\times (y_2-x_2+1)}$  denotes the sub-matrix  comprised of elements within the $x_1$-th to $y_1$-th rows and $x_2$-th to $y_2$-th columns of matrix $\mathbf{C}$.
Besides, define $\mathbf{q}=\left[\mathbf{q}_{1: k-1}^T, \mathbf{q}_{k: M}^T\right]^T$ and $\mathbf{d}=\left[\mathbf{d}_{1: k-1}^T, \mathbf{d}_{k: M}^T\right]^T$ where $\mathbf{q}_{x:y} $ and $\mathbf{d}_{x:y}  $  denotes the $(y-x+1)$-dimensional column vector comprised of elements from $x$-th  to $y$-th rows of $\mathbf{q}$ and $\mathbf{b}$, respectively. Then, condition (\ref{sphere_condition}) can be rewritten as
\begin{align}
	r^2 &   \ge  \|\mathbf{C q}-\mathbf{d}\|^2 = \left\|\mathbf{C}_{k: M, k: M} \mathbf{q}_{k: M}-\mathbf{d}_{k: M}\right\|^2\nonumber   \\ 
	&+\left\|\mathbf{C}_{1: k-1: 1: k-1} \mathbf{q}_{1: k-1}+\mathbf{C}_{1: k-1, k: M} \mathbf{q}_{k: M}
	- 
	\mathbf{d}_{1: k-1}\right\|^2 \nonumber  \\    \label{conventional_SB_condition}
	& \ge \left\|\mathbf{C}_{k: M, k: M} \mathbf{q}_{k: M}-\mathbf{d}_{k: M}\right\|^2,
\end{align}
which will be used by SD to prune the search tree of $\mathbf{q}$, starting from  $k=M$ and ending at  $k=2$.  Specifically, substituting $k=M$ into (\ref{conventional_SB_condition}), the bounding condition is $ \left|C_{M, M} q_M-d_M\right| \leq r $, leading to
\begin{align}\label{condition_M}
( -r+d_M) / C_{M, M}   \leq q_M \leq  (r+d_M)/  C_{M, M}    ,
\end{align}
 where $C_{a,b}$ is the $(a,b)$-th element of $\mathbf{C}$ and $q_a$ and $b_a$ denote the $a$-th entries of $\mathbf{q}$ and $\mathbf{b}$, respectively. (\ref{condition_M}) is the pruning condition at the  top level of the SD search tree, which enables us to examine the feasibility of $q_M \in \left\{-1,+1\right\}$. Given each one-dimensional point $q_M$ satisfying (\ref{condition_M}), we then move down to the next level with $k=M-1$, which clarifies the following pruning condition for $q_{M-1}$:
 \begin{align}\label{condition_M_minus_1}
\begin{aligned}
	& r^2 \geq\left\|\mathbf{C}_{M-1: M, M-1: M} \mathbf{q}_{M-1: M}-\mathbf{d}_{M-1: M}\right\|^2  \\
	& =\left(C_{M-1, M-1} q_{M-1}+    C_{M-1, M} q_M     -d_{M-1}\right)^2\\
	&+\left(C_{M, M} q_M-d_M\right)^2,
\end{aligned}
 \end{align}
and it can also be written   as an interval with respect to variable $q_{M-1}$     as (\ref{condition_M}). Given every node $\hat{q}_M$ satisfying (\ref{condition_M}), now (\ref{condition_M_minus_1}) determines the  feasibility of all branches $[\hat{q}_M, q_{M-1}]$. Keeping examining the feasibility of a branch from  $k=M$ to $k=2$ based on condition (\ref{conventional_SB_condition}), SD generates a search tree for $\mathbf{q}$ within which all points inside the sphere (\ref{sphere_condition}) are found. Then, we can obtain the globally optimal solution $\mathbf{q}^{\text{opt}}$ which achieves the minimal objective function.

However, as the number of antennas, i.e., $M$,  increases, the search space and computations of SD could increase dramatically. Thus,  it is important to improve the computation efficiency of SD in multi-cell HMIMO systems. To this end, we propose     exploiting the quadratic structure of    problem (\ref{SD_optimization_problem}) based on another necessary optimal condition\cite{beck2000global,wang2020low},  yielding the following lemma.
\begin{lem}\label{lemma_optimal_condition}
The optimal solution of the $k$-th element of $\mathbf{q}$ for problem (\ref{SD_optimization_problem}) can be decided if the following conditions meet:
\begin{align} \label{optimal_condition}
	q_k^{\mathrm{opt}}\! =\! \left\{   \!
	\begin{array}{l}
		+1, \text { if } \sum_{m=1, m \neq k}^M\left|\left[ \mathbf{A}^{\prime}      \right]_{k, m}\right| -  \left[\mathbf{b}\right]_k<0, \\
		-1, \text { if } \sum_{m=1, m \neq k}^M-\left|\left[    \mathbf{A}^{\prime}     \right]_{k, m}\right|  -  \left[     \mathbf{b}    \right]_k>0.
	\end{array}\right.
\end{align}
\end{lem}

\itshape Proof:
\upshape
Following \cite{beck2000global}, 	if $  \mathbf{q}^{\star}  =[q^{\star}_1,\cdots, q^{\star}_M]^T$ is a globally optimal solution of quadratic problem $  \left\|\mathbf{C} \mathbf{q}-\mathbf{d}\right\|^2 $ subject to $|[\mathbf{q}]_m|^2=1$, $\forall m$, then, $ \mathbf{q}^{\star} $ must satisfy
\begin{align}\label{dsfsfsg}
\operatorname{diag}\!\{\!\mathbf{q}\!^{\star}\!  \} \mathbf{C}^T \mathbf{C} \operatorname{diag}\!\{\! \mathbf{q}\!^{\star} \! \} \mathbf{1} \!-\!
\operatorname{diag} \! \{ \!  \mathbf{q}\!^{\star}\!\} \mathbf{C}^T \! \mathbf{d} \! \leq \! \operatorname{diag} \! \left\{ \! \mathbf{C}^T \mathbf{C} \! \right\} \! \mathbf{1},
\end{align}
where $\mathbf{1}\triangleq[1,\cdots, 1]^T$.  The necessary condition for $ [\mathbf{q}^{\star} ]_k$ can be extracted from  (\ref{dsfsfsg}) as  $q^{\star}_k   \kappa_k \leq 0 $, $k\in [1,M]$, where
\begin{align}
\kappa_k \triangleq  \sum\nolimits_{m=1, m \neq k}^M\left[\mathbf{C}^T \mathbf{C}\right]_{k, m} q^{\star}_m -
\left[\mathbf{C}^T \mathbf{d}\right]_k.
\end{align}
 Recall that    $  \mathbf{C}^T \mathbf{C} =  \mathbf{A}^{\prime} $ and $\mathbf{C}^T \mathbf{d}=  \mathbf{b} $.    Since $q^{\star}_k\in \{-1, +1\}$, $\forall k$, we have $\min \{{\kappa_k}\} \leq \kappa_k \leq \max \{{\kappa_k}\}$ where
 \begin{align}
\begin{aligned}
	& \max \{{\kappa_k}\}=\sum\nolimits_{m=1, m \neq k}^M\left|\left[   \mathbf{A}^{\prime}     \right]_{k, m}\right|  - \left[  \mathbf{b}\right]_k ,\\
	& \min \{{\kappa_k}\}=\sum\nolimits_{m=1, m \neq k}^M-\left|\left[  \mathbf{A}^{\prime}     \right]_{k, m}\right| - \left[  \mathbf{b}\right]_k.
\end{aligned}
 \end{align}
Therefore, to ensure $ q^{\star}_k   \kappa_k \leq 0  $, the optimal $q^{\star}_k$ must be positive  if $\max\{\kappa_k\}<0$. Also, if $\min\{\kappa_k\}>0$, the optimal $q^{\star}_k$ must be negative, which completes the proof. $\hfill \blacksquare$

Aided by Lemma \ref{lemma_optimal_condition}, before   conducting SD, we can firstly employ the optimal condition (\ref{optimal_condition}) for every element of  variable $\mathbf{q}$ and decide the possible optimal solutions in advance. Then, in the following SD operations, the levels associated with these pre-decided elements in the search tree are  no longer needed to be visited, which helps reduce the dimension of the SD problem.

In addition to applying the optimal condition (\ref{optimal_condition}), we propose incorporating HC algorithms to further speed up the SD. This can be achieved  by tightening the last lower bound in (\ref{conventional_SB_condition}) and pruning more branches. Specifically, for ${\mathbf{q}_{1: k-1} \in\{-1,+1\}^{k-1}} $, define the following lower bound  
\begin{align}\label{lower_bound}
\begin{aligned}
&\mathscr{R} (\mathbf{q}_{1: k-1})  \\
& \triangleq 
\left\|\mathbf{C}_{1: k-1,1: k-1} \mathbf{q}_{1: k-1}+\mathbf{C}_{1: k-1, k: M} \mathbf{q}_{k: M} - 
\mathbf{d}_{1: k-1}\right\|^2 \\
&\geq \operatorname{LB}_k\left(\mathbf{q}_{k: M}, \mathbf{C}_{1: k-1,1: M}, \mathbf{d}_{1: k-1}\right).
\end{aligned}
\end{align}
Then, the pruning condition (\ref{conventional_SB_condition}) can be tightened, for $1\leq k \leq M$, as
\begin{align}
\begin{aligned}
&r^2-\mathrm{LB}_k\left(\mathbf{q}_{k: M}, \mathbf{C}_{1: k-1,1: M}, \mathbf{d}_{1: k-1}\right) \\ &\geq\left\|\mathbf{C}_{k: M, k: M} \mathbf{q}_{k: M} -  \mathbf{d}_{k: M}\right\|^2.
\end{aligned}
\end{align}
Thus, the key task is to obtain a tight and low-complexity lower bound in (\ref{lower_bound}). Specifically, by relaxing the constraints, we have
\begin{align}\label{tighen_lowe_bound}
\mathscr{R}\left(\mathbf{q}_{1: k-1}\right) & \geq
 \min _{  \mathbf{q}_{1: k-1} =\{-1,+1\}^{k-1}  } 
 \mathscr{R}\left(\mathbf{q}_{1: k-1}\right) \\  \label{proBlemLB}
&\!  \!     \geq 
 \min _{ \left\|\mathbf{q}_{1: k-1}\right\|^2=k-1}
 \mathscr{R}   \left(\mathbf{q}_{1: k-1}\right) \triangleq \mathscr{R}   \left(   \mathbf{q}_{1: k-1}^{  {rel*}}     \right) .
\end{align}
By ignoring the  terms irrelevant to $\mathbf{q}_{1: k-1}$ in function $\mathscr{R}\left( \cdot\right) $, we can   find $   \mathbf{q}_{1: k-1}^{  {rel*}}    $ in  (\ref{proBlemLB}) as follows
\begin{align}\label{lower_bound_SD}
	\begin{aligned}
	\mathbf{q}_{1: k-1}^{  {rel*}}  \! \!	&=  \!\! \!    \underset{\left\|\mathbf{q}_{1: k-1}\right\|^2=k-1}{\arg \min} \mathbf{q}_{1: k-1}^T \mathbf{C}_{1: k-1,1: k-1}^T \mathbf{C}_{1: k-1,1: k-1} \mathbf{q}_{1: k-1} \\
		& -2\left(    \mathbf{d}_{1: k-1}   -\mathbf{C}_{1: k-1, k: M} \mathbf{q}_{k: M}    \right)^T \mathbf{C}_{1: k-1, 1: k-1} \mathbf{q}_{1: k-1},
	\end{aligned}
\end{align} 
which has the same structure as problem (\ref{problem_relax_quadratic}) and therefore   can be effectively tackled  by employing the HC-based algorithm proposed in the previous subsection.

Strictly speaking, the SD algorithm possesses the complexity of $\mathcal{O}(2^{ {\eta}M})$ where $\eta \in (0,1]$ is some small factors depending on SNR and other system parameters\cite{jalden2005complexity}. Nevertheless, in practice, it could be $\eta \ll 1 $  and the complexity of SD could be dominated by polynomial terms.  Meanwhile, with the help of the optimal condition (\ref{optimal_condition}), the number of levels in the search tree can be reduced from $M$ to $M-M_{\rm dec}$ where $M_{\rm dec}$ is the number of elements in $\mathbf{q}$ that can be decided from condition (\ref{optimal_condition}). Furthermore, after tightening the sphere radius with the obtained lower bound, much more branches can be pruned from the search tree, at the cost of increased computations   in each level. There is a trade-off between managing the computational complexity required to calculate the lower bound and expanding the scale of the pruned tree in SD.  Fortunately, for every level $k-1 \in [1,M-1]$, most of the parameters required  to calculate the HC-based lower bound from (\ref{lower_bound_SD}) can be calculated in advance before conducting SD. Therefore,  the additional complexity when incorporating the SD algorithm with HC-based  lower bound is   $\mathcal{O}((k-1)^3)$ for each level $k-1$.  Overall, the complexity of the proposed accelerated SD could be $\mathcal{O}(2^{ {\eta}(M - M_{\rm dec})} + M^4)$.

Finally, the pseudocode of the revised SD algorithm is presented in Algorithm \ref{algorithmSD}, where set $\mathcal{K}^o$ stores the nodes that can be decided optimally  based on Lemma \ref{lemma_optimal_condition} and therefore the corresponding levels in the search tree do not need to be visited in the entire algorithm. To reduce the computations, as shown in  relationship (\ref{condition_M_minus_1}), two parameters of SD are calculated recursively and defined as follows
\begin{align}\label{update_criteria}
	&	\mathrm{g}_k^{>}  \triangleq  [\mathbf{C q}-\mathbf{d}]_k-C_{k, k} q_k \!=\mathbf{C}_{k: k, k+1: M} \mathbf{q}_{k+1: M}-d_k , \\ 
	&	\left(  \widehat{r}^k  \right)^2  =\left(\widehat{r}^{k+1}\right)^2-\left(\mathrm{g}_{k+1}^{>}+C_{k+1, k+1} q_{k+1}\right)^2.
\end{align}
 Step \ref{bounding} is the basic bounding operation and step \ref{bouding2} is the tighter bounding based on the lower bound from (\ref{lower_bound_SD}). $k=1$ indicates that the search reaches the bottom of the tree and thus we conduct the possible update of the solution, radius, and radius difference $\nabla$  in step \ref{solution}.

\begin{algorithm}[t]
	\caption{Accelerated SD algorithm.}
	\begin{algorithmic}[1]\label{algorithmSD}
		\REQUIRE Parameters $\mathbf{C}$ and $\mathbf{d}$ of problem (\ref{SD_optimization_problem}), initial radius $r^\star$,   $\mathcal{K}^o=\left\{k| q_k^{\text{opt}} \text{can be decided in (\ref{optimal_condition})}, 1\leq k \leq M\right\}$;
		\ENSURE $ \mathbf{q}^\star $  
		\STATE Initialize $\mathbf{q}=\{-1\}^M$. Let $  [\mathbf{q}]_k = q_k^{\text{opt}}$, $\forall k \in \mathcal{K}^o$. $r=r^{\star}$, $\nabla=0$, $\widehat{r}^M=r$, $\mathrm{g}_M^{>}=-d_M$, $k= M$;
		\STATE \textbf{While} {$k\in \mathcal{K}^o$}, \textbf{do} $k=k-1$ and update $ g_k^{>} $ and $ \widehat{r}^k  $ based on (\ref{update_criteria}) when $k> 1$. Then, if $k=0$, terminate;
		\STATE (Basic bounding) $ \overline{\mathrm{lb} }_k    = (-\sqrt{ (\widehat{r}^k )^2-    \nabla    }-\mathrm{g}_k^{>}) / C_{k, k}$ and $\overline{\mathrm{ub}}_k = ( \sqrt{ (\widehat{r}^k )^2-  \nabla    }-\mathrm{g}_k^{>} ) / C_{k, k}$; \label{bounding_obtain}
		\IF {$  \overline{\mathrm{lb} }_k      \le q_k \le     \overline{\mathrm{ub} }_k$,} \label{bounding}
					\IF{$k=1$ or $(\widehat{r}^k)^2- \nabla - (\mathrm{g}_k^>+C_{k, k} q_k )^2 \geq 
						\operatorname{LB}_k\left(\mathbf{q}_{k: M}, \mathbf{C}_{1: k-1,1: M}, \mathbf{d}_{1: k-1}\right)$,} \label{bouding2}
									\STATE  Go to step \ref{move_down};
					\ELSE 
					 \STATE Go to step \ref{next_node};
					\ENDIF
		\ELSE
				     \STATE Go to step \ref{next_node};
		\ENDIF
		\STATE (Next node)  if  $  [\mathbf{q}]_k=-1 $, $  [\mathbf{q}]_k=1 $ and go to step \ref{bounding}; else, go to step \ref{move_up};   \label{next_node}
		\STATE (Move up) \textbf{while} {$k+1\in \mathcal{K}^o$}, \textbf{do} $k=k+1$. Then,  if $k=M$, terminate; else, $k=k+1$ and go to step \ref{next_node};  \label{move_up}
		\STATE (Move down) \textbf{while} {$k-1\in\mathcal{K}^o$}, \textbf{do}  $k=k-1$ and update $ \mathrm{g}_k^{>} $ and $ \widehat{r}^k  $ based on (\ref{update_criteria}). Then, if $k=1$, go to step \ref{solution}; else, $k=k-1$,  update $ \mathrm{g}_k^{>} $ and $ \widehat{r}^k  $ as (\ref{update_criteria}), $ [\mathbf{q}]_k=-1 $, and go to step \ref{bounding_obtain}; \label{move_down}
		\STATE (Solution update) if $  \|\mathbf{C q} - \mathbf{d}\| <r^\star$, update $r^\star= \|\mathbf{C q}- \mathbf{d}\|$, $\nabla=r^2-(r^\star)^2$, and $\mathbf{q}^\star= \mathbf{q}$. Then, go to step \ref{next_node};  \label{solution}
	\end{algorithmic}
\end{algorithm}

\subsection{Design with Grayscale Constraints}
In this section, we propose an MM-based algorithm to tackle the HMIMO tunable response design under the Lorentzian-constrained phase model in (\ref{gray_cons}). Recall optimization problem (\ref{optimization_HMIMO_phase}) with constraint set $\mathcal{Q}^g$. To begin with, denoting $\mathbf{q}_{i^{\prime}}=( j \mathbf{1}+\hat{\mathbf{q}}_{i^{\prime}}) /2  $ and substituting it into (\ref{30a}), the optimization variable of  problem (\ref{optimization_HMIMO_phase}) can be transferred from Lorentzian-constrained variables $ \mathbf{q}_{i^{\prime}} $ to  unit-modulus variables  $\hat{\mathbf{q}}_{i^{\prime}}$. Then, by ignoring the constant terms,  optimization problem with respect to $ \hat{\mathbf{q}}_{i^{\prime} }$ can be formulated as
\begin{subequations}\label{mm_problem}
	\begin{align} \label{mm_funtion}
	& \min _{ \hat{\mathbf{q}  }_{i^\prime}} f_{\hat{u}}\left(\hat{\mathbf{q}}_{i^{\prime}}\right)=\frac{1}{4} \hat{\mathbf{q}}_{i^{\prime}}^H \mathbf{A}_o \hat{\mathbf{q}}_{i^\prime}
	- \operatorname{Re}\left\{\left(\mathbf{b}_o^T + \frac{j}{2} \mathbf{1}^H \mathbf{A}_o\right) \hat{\mathbf{q}}_{i^{\prime}}\right\} \\\label{unite_modulus_constr}
& \text { s.t. } \quad \left|\left[\hat{\mathbf{q}}_{i^{\prime}}\right]_m \right|=1, \forall m.
	\end{align}
\end{subequations}	
This problem is  non-convex due to the non-convex unit-modulus constraint. It can be tackled effectively based on MM algorithms. To this end, we need to find a tractable upper-bounded surrogate  function for $  f_{\hat{u}}\left(\hat{\mathbf{q}}_{i^{\prime}}\right) $ in (\ref{mm_funtion}).

Specifically, at a fixed point $ \bar{\mathbf{q}}_{0}$, we have the following inequality for the quadratic term\cite[(61)]{zhi2022ZF}
\begin{align}\label{MM_bound}
\hat{\mathbf{q}}_{i^{\prime}}{ }^H \mathbf{A}_o \hat{\mathbf{q}}_{i^{\prime}} \leq -2 \operatorname{Re}\left\{\hat{\mathbf{q}}_{i^{\prime}}{ }^H\left(
\lambda_{\max }\left\{\mathbf{A}_o\right\} \mathbf{I}_M  -\mathbf{A}_o   
\right) \overline{\mathbf{q}}_0
\right\}+c_{\overline{\mathbf{q}}_0},
\end{align}
 where
 \begin{align}
c_{\overline{\mathbf{q}}_o}=M \lambda_{\max }\left\{\mathbf{A}_o\right\}+\overline{\mathbf{q}}_0^H\left(\lambda_{\max }\left\{\mathbf{A}_o\right\} \mathbf{I}_M-\mathbf{A}_o\right) \overline{\mathbf{q}}_0,
 \end{align}
following the properties that $ \lambda_{\max}\left\{\mathbf{A}_o\right\} \mathbf{I}_M \succeq  \mathbf{A}_o $ and $\hat{\mathbf{q}}_{i^{\prime}}^H \lambda_{\text {max}}\left\{\mathbf{A}_o\right\} \mathbf{I}_M \hat{\mathbf{q}}_{i^{\prime}}=M \lambda_{\text {max}}\left\{\mathbf{A}_o\right\} $. The equality of ($\ref{MM_bound}$) holds when $\hat{\mathbf{q}}_{i^{\prime}} =  \bar{\mathbf{q}}_{0}$.

Substituting (\ref{MM_bound}) into (\ref{mm_funtion}), the objective function of (\ref{mm_problem}) is upper bounded by
\begin{align}\label{surrogate_dunction}
	f_{\hat{u}}\left(\hat{\mathbf{q}}_{i^{\prime}}\right) \leq -\operatorname{Re}\left\{\mathbf{z}^H \hat{\mathbf{q}}_{i^{\prime}}\right\}+c_{\overline{\mathbf{q}}_0}/4, 
\end{align}
where
\begin{align}
\mathbf{z}^H=\frac{1}{2} \overline{\mathbf{q}}_0{ }^H\left(
\lambda_{\max }\left\{\mathbf{A}_o\right\} \mathbf{I}_M
\! -\!  \mathbf{A}_o
\right)^H
+\mathbf{b}_o^T + \frac{j}{2} \mathbf{1}^H \mathbf{A}_o.
\end{align}
The surrogate function in (\ref{surrogate_dunction}) is linear and admits low-complexity solutions. Under the unit-modulus constraint (\ref{unite_modulus_constr}),  the closed-form optimal solution to minimize the surrogate function can be directly given by
\begin{align}\label{optimal_solution}
\underset{  \hat{\mathbf{q}}_{i^{\prime}}    }{\arg \min }\left\{-\operatorname{Re}\left\{\mathbf{z}^H \hat{\mathbf{q}}_{i^{\prime}}\right\}+c_{\overline{\mathbf{q}}_0} /4\right\}=\exp \{j \angle \mathbf{z}\},
\end{align}
where $[\mathbf{z} ]_m\triangleq |z_m| e^{j \angle z_m}$, $\forall m$, and $\angle \mathbf{z}=[\angle z_1, \cdots, \angle z_M]^T$. The solution obtained from (\ref{optimal_solution}) will be used to update the fixed point. With the new fixed point, we can re-derive the upper bound and obtain a new optimal solution. Repeating this process until convergence,   problem (\ref{mm_problem}) will  be  iteratively solved with solution $\hat{\mathbf{q}}_{i^{\prime}} ^{\text{opt}}$. Then, the optimized solution for problem (\ref{optimization_HMIMO_phase}) can be mapped back as $\mathbf{q}_{i^{\prime}}^{\text{opt}}=( j \mathbf{1}+\hat{\mathbf{q}}_{i^{\prime}}^{\text{opt}}) /2  $.

\subsection{Overall Algorithms}
\begin{algorithm}[t] 
	\caption{Overall algorithm.}
	\begin{algorithmic}[1]\label{algorithm2}
		\STATE Input $\omega_{{i_k}}$, $P_{i,m}$.  Initialize  $\mathbf{W}_{i_k}$ and $\mathbf{Q}_i$
		\REPEAT 
		\STATE Update $\mathbf{U}_{i_k} $ by (\ref{optimal_U}), $\forall i_k$;
		\STATE Update $\mathbf{V}_{i_k}$ by (\ref{optimal_V}), $\forall i_k$;
		\STATE Update $\tilde{\mathbf{w}}_{i,m}$ by (\ref{optimal_w}), $\forall i_k, m$;
		\IF{under binary constraints $\mathcal{Q}^b$}
		\STATE Update $\mathbf{Q}_i$  via (\ref{optimal_Q_hidden}) or  SD via Algorithm \ref{algorithmSD}, $\forall i$; 
		\ELSIF{under grayscale constraints $\mathcal{Q}^g$}
		\STATE Update $\mathbf{Q}_i$ by MM  algorithm (\ref{optimal_solution}) iteratively, $\forall i$;
		\ENDIF
		\UNTIL the objective value in (\ref{optimization_problem}) converges.
	\end{algorithmic}
\end{algorithm}
The description of overall algorithms with BCD is given in Algorithm \ref{algorithm2}. The algorithms are guaranteed to converge due to the monotonically decreasing values of the MMSE objective function in (\ref{MSE_problem}) and also due to the limited lower bound with the power constraint. Following the previous discussions in each subsection,  the overall computational complexities of the proposed algorithms per iteration are summarized in Table \ref{tab1}.
\begin{table}[t]
	\centering 
	\caption{Computational Complexity}	
	\begin{tabular}{|c|c|c|c|c|c|c}
		\hline
		WWSE-HC&$\mathcal{O}\{  L (         M^3    + L  \overline{U}  M^2 d  +    \overline{U}^2 M^2        )          \}$\\
		\hline
		WWSE-SD&  $\mathcal{O}\{   L (     2^{ {\eta}(M - M_{\rm dec})} +M^4 + L  \overline{U}  M^2 d     )    \} $\\
		\hline
		WWSE-MM& $ \mathcal{O}\{  L (      M^3    + L  \overline{U}  M^2 d  +    \overline{U}^2 M^2      ) \} $\\
		\hline
	\end{tabular}\label{tab1}
\end{table}

\section{SNR Analysis for MISO Case}  \label{section4}
In this section, we aim to explicitly analyze the performance of HMIMO under binary and grayscale constraints and compare it with conventional fully digital and hybrid D/A arrays. Our objective is to characterize SNR scaling laws and answer the fundamental question of whether HMIMO can achieve better performance or not compared to conventional arrays.

To facilitate the analysis,  we focus on the basic  single-cell ($L=1$) MISO downlink scenario where a single-antenna user is served by a linear HMIMO with a single microstrip ($M_y = 1$) consisting of $M$ metamaterial elements. Meanwhile, only the LoS channel is considered to exist. 
With a slight abuse of notations in (\ref{notationsss}), we denote  the propagation  within the waveguide by $\mathbf{t}$ where $[ \mathbf{t} ]_m =  e^{-   \Delta (m-1)  (\alpha + j\beta) } $ and the LoS channel from linear HMIMO array to the user by $ \mathbf{a}_M(\phi)$ where $ [ \mathbf{a}_M(\phi)]_m =   e^{j \frac{2\pi \Delta}{\lambda}(m-1) \sin (\phi)}  $, respectively. 	Denote $q_m=\left[\mathbf{Q}\right]_{m m}=\left[\operatorname{diag}\left\{\mathbf{q}\right\}\right]_{m m}$ as HMIMO response or conventional array's analog precoder   at antenna $m$. 

Then,  given the same total power budget $p$, the received signal at the user from the linear $M$-antenna digital array, sub-connected hybrid D/A array, and HMIMO with grayscale  and binary constraints can be expressed, respectively, as\footnote{Compared to conventional antennas, a larger number of HMIMO metamaterial antenna elements can be deployed at the given array aperture  due to the smaller element size and spacing.  However,  actual receiving power is related to the antenna radiating area and therefore, a proper normalization must be introduced. Following \cite{tang2020wireless,zhi2024performance}, to obtain results with physical significance, we add a normalization factor $\sqrt{A}$ to model the impact of antenna area.}
\begin{align}\label{snr_4}
\begin{aligned}
	& y_D=\sqrt{\!A}\mathbf{a}_M^H(\phi) \mathbf{w} s + \tilde{n}, \quad\quad  \;\;  \|  \mathbf{w}  \|^2 \leq p,\\
	& y_{D / A}=\sqrt{\!A}\mathbf{a}_M^H(\phi) \mathbf{q}  \mathrm{w} s + \tilde{n}, \quad  \left|q_m\right|=1, \|   \mathbf{q}   \mathrm{w} \|^2 \leq p  , \\
	& y_g \!=\! \sqrt{\!A}\mathbf{a}_M^H(\phi) \mathbf{Q} \mathbf{t } \mathrm{w} s \!+\!    \tilde{n}, \; q_m \! \in \frac{e^{j \theta}+j}{2}
	,\| \mathbf{Q} \mathbf{t } \mathrm{w}\|^2 \leq p, \\
	& y_b=\sqrt{\!A}  \mathbf{a}_M^H(\phi) \mathbf{Q} \mathbf{t} \mathrm{w}s + \tilde{n}, \; q_m \! \in\{0,1\},\|\mathbf{Q} \mathbf{t }  \mathrm{w}\|^2 \leq p,
\end{aligned}
\end{align}
with   effective radiating area $A$ for each antenna element, symbol $\mathbb{E}\{|s|^2\}= 1$  and  noise $\tilde{n} \sim \mathcal{C N}(0,\sigma^2)$.

 
\begin{theorem}\label{SNR_theorem}
Define  $\psi=\frac{2\pi \Delta}{\lambda} \sin (\phi)+\Delta\beta  $ and indicator function $ \mathbb{I}(\cdot)$ where $ \mathbb{I}(x)=1$, if $x>0$ ; $\mathbb{I}(x)=0$, if $x \leq 0$.  The receiving SNR realized by digital array, hybrid D/A array, HMIMO with grayscale constraint, and HMIMO with binary constraint   are, respectively, given by
\begin{align}\label{Digital}
	& \operatorname{SNR}_D=\frac{p}{\sigma^2} MA , \quad\quad \mathbf{w} =\sqrt{p/M} \; \mathbf{a}_M(\phi), \\\label{HybridDA}
	& \operatorname{SNR}_{D / A}=\frac{p}{\sigma^2} MA ,  \quad \mathbf{q}=\mathbf{a}_M (\phi), \;\;  \mathrm{w}=   \sqrt{p/M} ,\\\label{HMIMOGrey}
	& \operatorname{SNR}_g \approx \frac{p }{2 \sigma^2} MA , \;\;  q_m =   \frac{e^{j(m-1) \psi}+j}{2} , \mathrm{w}=     \sqrt{ \frac{2p}{M} }  ,\\\label{HMIMOBI}
	& \operatorname{SNR}_b \approx \frac{2p}{\pi^2 \sigma^2} MA,  q_m \! = \!   \mathbb{I}   (\cos ((m \!-\!  1) \psi)), \mathrm{w} \!= \! \! \sqrt{  \!   \frac{2p}{M}}.
\end{align}
\end{theorem}

\itshape Proof: \upshape  (\ref{Digital}) and (\ref{HybridDA}) can be proved readily and  omitted here. Next,  due to the small value of $\Delta $ which is on the order of sub-wavelength and $\alpha<1$, for a medium value of $M$, we have $\mathbf{t} \approx \widetilde{\mathbf{t}}$ where $ [ \widetilde{\mathbf{t}}  ]_m = e^{-  j  (m-1)   \Delta \beta  } $. Then, with  design of $q_m$ in (\ref{HMIMOGrey}), we tackle the constraint of $\mathrm{w}$ as follows
\begin{align}
	& \left\|\mathbf{Q}_M \mathbf{t }\mathrm{w}\right\|^2 \approx\left\|\mathbf{Q}_M \tilde{\mathbf{t}}\mathrm{w}  \right\|^2=|\mathrm{w}|^2 \sum\nolimits_{m=1}^M\left|q_m\right|^2 \nonumber\\
	& =\frac{|\mathrm{w}|^2 }{4}   \!  \! \sum_{m=1}^M \! \!   \left\{   \!  2   \!  -   \!   j e^{j(m-1) \psi}   \!  + \!     j  e^{-j(m-1) \psi}\right\} \stackrel{(a)}{\approx} \frac{|\mathrm{w}|^2}{2} M,  
\end{align}
where $(a)$ used $ \sum_{m=1}^M e^{-j(m-1) \psi}=\frac{\sin (M \psi / 2)}{\sin (\psi / 2)} e^{-j(M-1) \psi / 2} $ \cite[(153)]{zhiTwotimescale2022}. Therefore, we design $  \mathrm{w}\!=\!\sqrt{2p/M}  $  in (\ref{HMIMOGrey}) and the SNR is approximated as
\begin{align}
	&  \operatorname{SNR}_g \!  \approx   \!   \frac{2Ap}{M\sigma^2}\left|\mathbf{a}_M^H(\phi) \mathbf{Q} \tilde{\mathbf{t}}\right|^2  
	  \! =   \! \frac{2Ap}{M\sigma^2}    \! \left|    \! \frac{M}{2}   \! +   \!    \frac{j}{2} \sum_{m=1}^M      \!     e^{-j(m-1) \psi}\right|^2 \nonumber\\
	&=  \!  \frac{2Ap}{M\sigma^2}\left|     \frac{M}{2}    \! +  \!  \frac{\sin (M \psi/2)}{2 \sin (\psi/2)} e^{-j \frac{(M-1) \psi-\pi}{2}}\right|^2    \!  \approx \frac{p}{2 \sigma^2} MA.
\end{align}
 In (\ref{HMIMOBI}), as a projection of optimal solution $e^{j(m-1) \psi}$ onto  $\{0,1\}$, $q_m$ is designed in a heuristic way of $  q_m=  \mathbb{I}   (    \mathrm{Re} \{ e^{j(m-1) \psi}  \}     ) $. $\mathrm{w}$ is designed the same as (\ref{HMIMOGrey}) since the binary case is a special case of greyscale case.
 Then, we have
\begin{align}\label{fwefw}
	& \operatorname{SNR}_b \approx \frac{2Ap}{M\sigma^2}\left|\sum\nolimits_{m=1}^M e^{-j(m-1) \psi}  \mathbb{I}\{\cos ((m-1) \psi)\}        \right|^2 \nonumber \\
	& \stackrel{(b)}{=}  \frac{2Ap}{M\sigma^2}  
	\bigg |\sum\nolimits_{m=1}^M e^{-j(m-1) \psi}   \Big \{        \frac{1}{2}+\sum\nolimits_{n=1}^{\infty} \frac{1}{\pi n} \sin \left(\frac{\pi}{2} n\right)     \nonumber \\
	&     \times  \Big(   e^{j n(m-1) \psi}+e^{-j n(m-1) \psi}    \Big) \Big\}       \bigg|^2,
\end{align}
where $(b)$ applies the Fourier series expansion of even-periodic function $\mathbb{I}\{\cos ((m-1) \psi)\}$ with a period of $ 2\pi/(m-1) $ and employs $\cos ((m-1) n\psi)= (e^{j n(m-1) \psi}+e^{-j n(m-1) \psi} )/2$. Then, by using \cite[(153)]{zhiTwotimescale2022}, after some algebraic manipulations, we can approximate the SNR in (\ref{fwefw}) by the dominant term as follows
\begin{align}
\operatorname{SNR}_b \stackrel{n=1}{\approx} \frac{2 A p}{\sigma^2 M}\left|  
\sum_{m=1}^M e^{-j(m-1) \psi} \frac{e^{j(m-1) \psi}}{\pi}
\right|^2=\frac{2 p}{\pi^2 \sigma^2} M A,
\end{align}
which completes the proof.
$ \hfill \blacksquare  $

\begin{figure}[t]
	\centering
	\includegraphics[width= 0.4\textwidth]{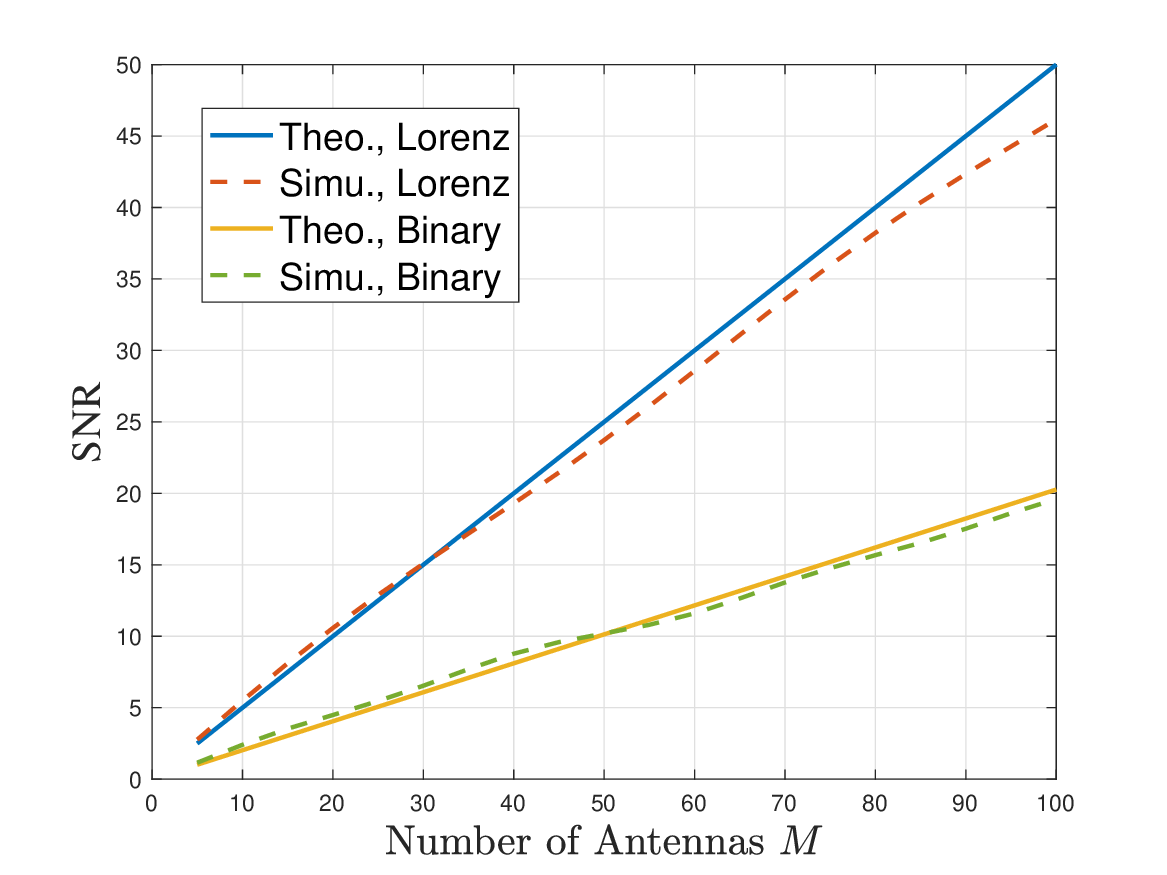}
	\DeclareGraphicsExtensions.
	\caption{SNR performance for holographic MISO systems, with $f=28$ GHz, $\Delta = \mathrm{c}/(8 f)$, $p/ \sigma^2 = 1$, $A=1$, $\alpha=0.6$, $\beta=29.56f/10^9$\cite{zhang2022beam,xu2024near}.}
	\label{figureSNR}
\end{figure}
We first verify the accuracy of our theoretical derivations in Fig. \ref{figureSNR}. It can be seen that the theoretical results based on (\ref{HMIMOGrey}) and (\ref{HMIMOBI}) are in very good agreement   with simulation results for   $M<100$. Thus,  in the following, we  draw insights on the performance of HMIMO based on Theorem \ref{SNR_theorem}. 

It can be seen that even though with simple circuits,  SNRs of HMIMO still  scale linearly with $M$ as conventional fully digital and hybrid D/A arrays, but with slightly lower coefficients due to reduced beam quality caused by limited adjustable freedom of $\mathbf{q}$. 
Note that given an array aperture, HMIMO can accommodate more antenna elements (i.e.,  larger  $ M $) than conventional arrays, due to smaller element size (i.e., smaller $A$) and spacing.  As a result, given the same overall array area (approximately $ MA $), HMIMO may not provide higher SNRs compared to conventional arrays, as characterized in Theorem \ref{SNR_theorem}.
Nevertheless, with simple diode-based circuits, HMIMO has strong advantages in hardware implementation, power consumption, and cost. In particular, the series feeding feature of  HMIMO (i.e., only one feeding line per microstrip) leads to a much simpler wiring layout compared with the parallel feeding mode of phased arrays where  feeding lines are required by all antenna elements\cite{deng2023reconfigurable}. Thus, it is circuit-feasible and cost-efficient to implement   extremely large-scale MIMO (i.e., with a large product $ M A $) based on HMIMO structures in future industry deployment,     bringing superior performance in terms of data rate, power saving, and cost reduction. 


\section{Simulation Results}\label{section5}

In this section, numerical results are presented to validate the effectiveness of the proposed algorithms and evaluate the performance of the multi-cell HMIMO systems. Unless statement otherwise, we consider  $L=3$ cell with an inter-cell distance of $400$ m, with $U_i=2$ users in each cell and $d=2$ streams for each user. The number of antennas on each user and BS is $N=2$ and $M=32$, respectively. The number of RF chains is $M_y = 4$ and number of antennas on each microstrip is $M_x=8$. The maximal power at each RF chain is $P_{i,m} = P_{\mathrm{tot}}/M$, $\forall i, m$, with the sum-antenna power budget $P_{\mathrm{tot}} = 30$ dBm. The variance of noise is $\sigma^2= -104$ dBm. We consider the mmWave frequency band of $f=28$ GHz and  adopt the mmWave channel model as  \cite{fang2021hybrid}. The propagation factors of waveguide are $\alpha_{i,{m_y}} = 0.6$, $\beta_{i,{m_y}} = 29.56f/10^9$\cite{zhang2022beam,xu2024near}, and $\varpi = 0.8$. The element spacing on the microstrip is $\lambda/8$ and the spacing between microstrips is $\lambda/2$.   

Some  schemes and benchmarks are summarized below:
 \begin{itemize}
\item  \textbf{WMMSE-HC/SD/BiProj}: design  baseband precoder  as (\ref{optimal_w})  and design binary tuning  response of HMIMO   by     HC/SD/closest point projection (\ref{simple_binary}).

\item \textbf{WMMSE-MM/GrayProj}: design   baseband precoder  as (\ref{optimal_w})   and design grayscale tuning  response of HMIMO   by    MM algorithm/closest point projection  (\ref{simple_gray}).

\item \textbf{ZF-HC}: design  baseband precoder by using CVX toolbox to find intra-cell zero-forcing solutions with per-RF power constraints and design binary tuning  response of HMIMO    by    HC algorithm.

\item \textbf{Fully Digital Array}: design the baseband precoder  as (\ref{optimal_w})  with $\mathbf{Q}_i = \mathbf{I}_M$ and $\mathbf{T}_i = \mathbf{I}_M$.

\item \textbf{Sub-Connected Hybrid D/A Array}: design   baseband precoder as (\ref{optimal_w})  and design  unit-modulus analog precoder     by slightly modifying the proposed MM algorithm and substituting non-zero elements in $ \mathbf{T}_i$ with $1$.
 \end{itemize}

\begin{figure}[t]
	\centering
	\includegraphics[width= 0.4\textwidth]{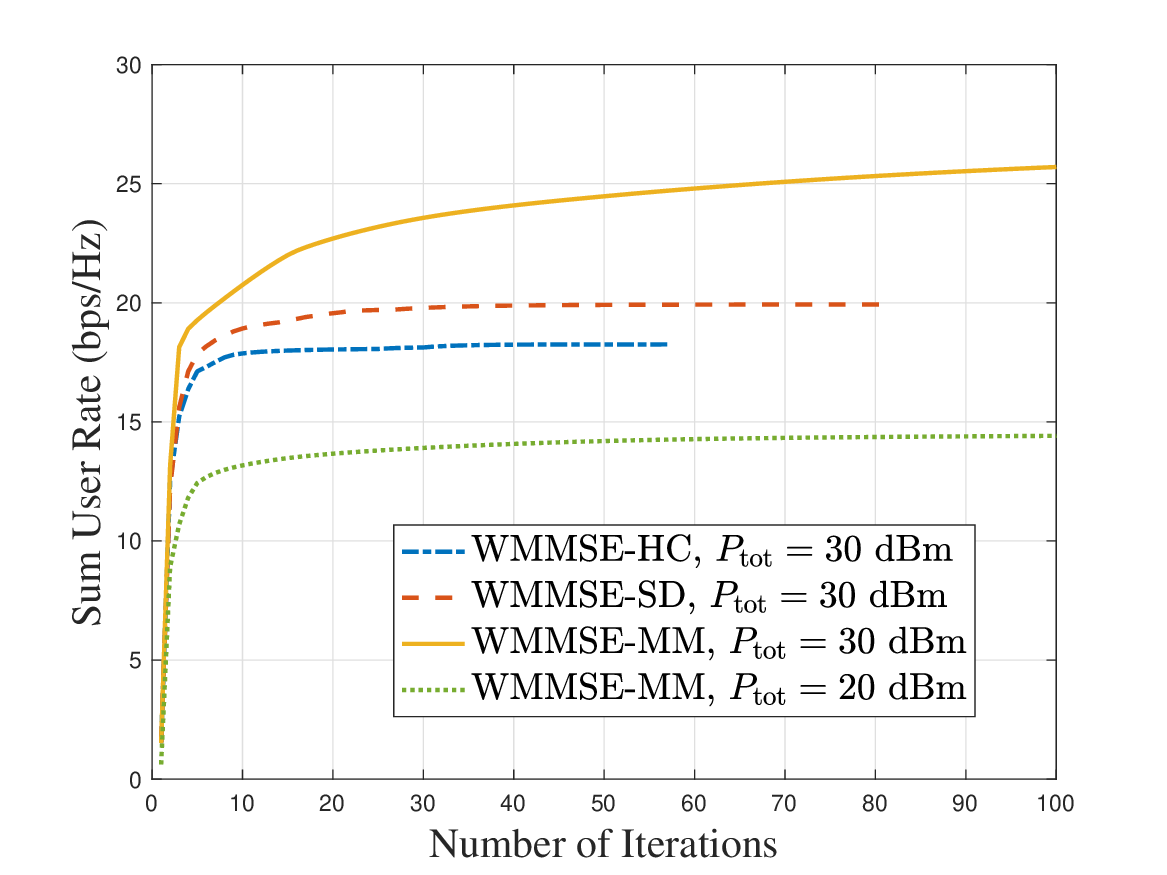}
	\DeclareGraphicsExtensions.
	\caption{Convergence behavior of the proposed algorithm.}
	\label{figure3}
\end{figure}

Fig. \ref{figure3} illustrates the convergence behavior of the proposed  algorithms. It can be seen that for the considered multi-cell networks with per-RF chain power constraints, the proposed algorithm can converge quickly. Meanwhile, it can be seen that the sum user rate monotonically increases with the iterations. 
Besides, it can be seen that HMIMO design under binary tuning constraints converges quicker than that under grayscale tuning constraints. Furthermore, the number of iterations needed in the algorithm slightly increases with the power budget for RF chains, due to the larger design freedom for precoding variables and more complex features of  intra-cell and inter-cell interference.

\begin{figure}[t]
	\centering
	\includegraphics[width= 0.4\textwidth]{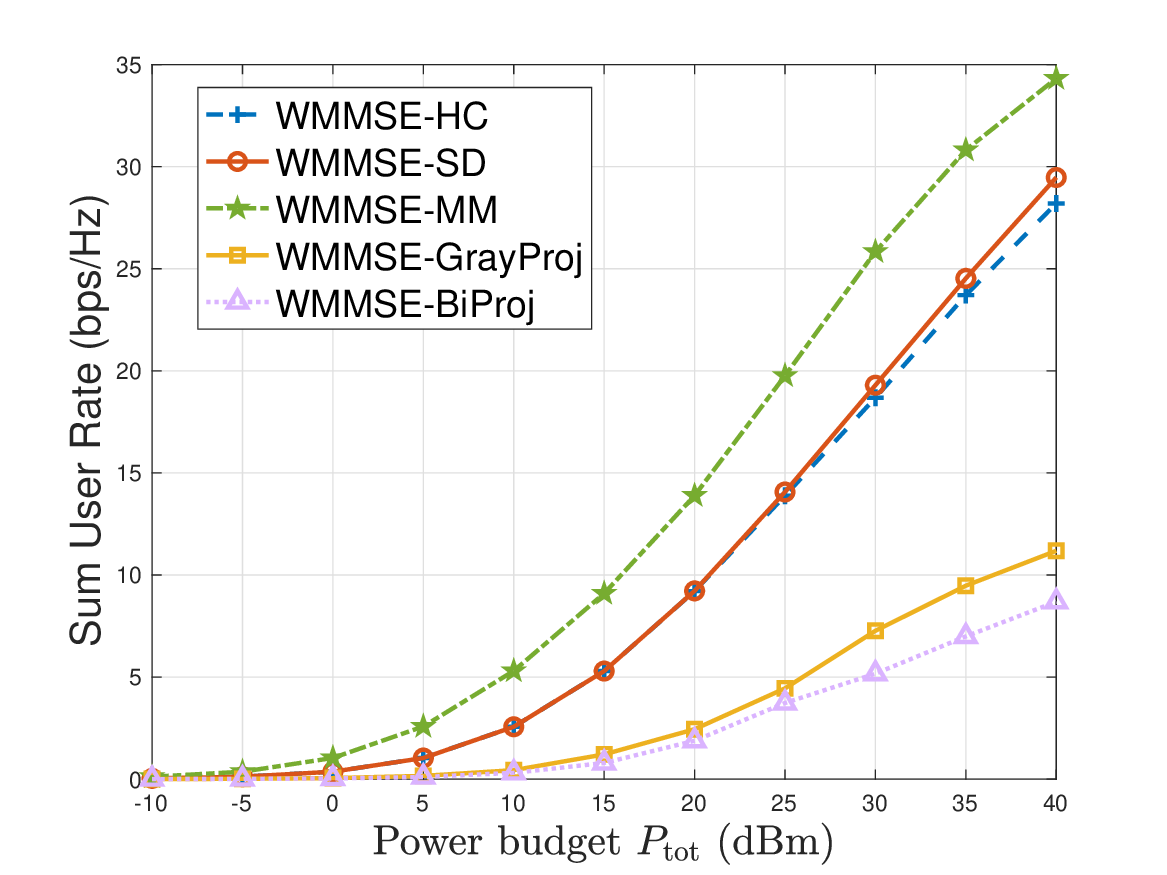}
	\DeclareGraphicsExtensions.
	\caption{Sum user rate versus the total power budget where $ P_{i,m} = P_{\mathrm{tot}}/M $.}
	\label{figure4}
\end{figure}

In Fig. \ref{figure4}, we evaluate the performance of the sum user rate with the increased power budget for   RF chains. Each curve in the sequel of this section is averaged over $100$ random channel realizations and user distributions. It can be seen that under the grayscale tuning constraint (i.e., WMMSE-MM), HMIMO can achieve a better performance than that with binary tuning constraints. This is reasonable, since binary tuning response can be viewed as a one-bit quantization of Lorentzian phase constraints and therefore sacrificing some beamforming precision. This result also agrees with our derivations in Theorem \ref{SNR_theorem} where $\mathrm{SNR}_g$ is higher than $\mathrm{SNR}_b$.
Nevertheless, binary-constrained HMIMO with PIN diodes owns a   simpler hardware complexity and implementation cost and therefore it is promising for practical large-scale deployment. Besides, it can be seen that the SD algorithm achieves slightly better performance than the HC algorithm and their gap slowly increases with the power budget.  This is because  the SD algorithm can find globally optimal solutions in each iteration while the HC algorithm can be sub-optimal. However, SD could result in much higher complexities than HC.  Thus, we can conclude   that due to the closed-form solution and tighter relaxation feature of the proposed HC algorithm, it achieves a great trade-off between complexity and performance.  Furthermore, it can be seen that the closest point projection methods, both in binary and grayscale cases, produce low sum user rates, which demonstrates the effectiveness of the proposed algorithms.

\begin{figure}[t]
	\centering
	\includegraphics[width= 0.4\textwidth]{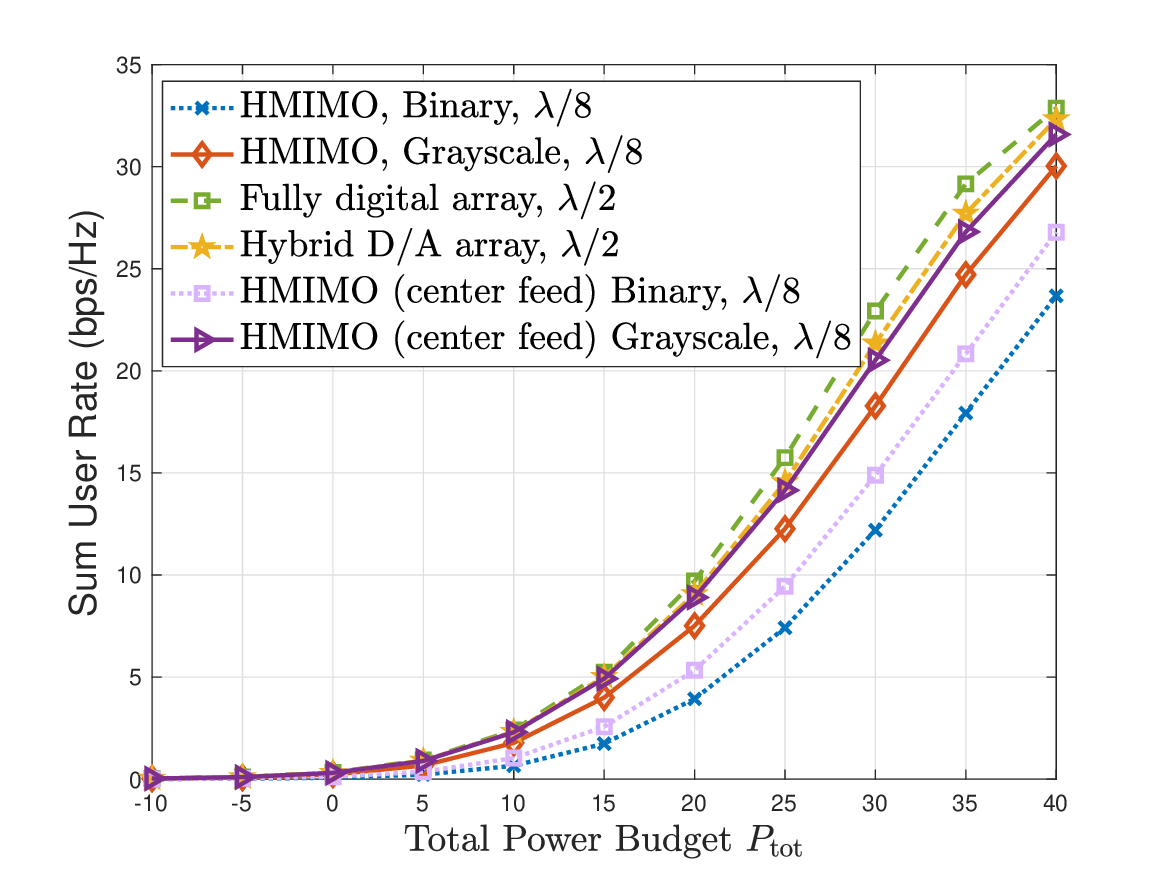}
	\DeclareGraphicsExtensions.
	\caption{Comparison between HMIMO and conventional arrays, $M_y=10$.}
	\label{figure6}
\end{figure}

Fig. \ref{figure6} compares the HMIMO structure with conventional fully digital and hybrid D/A arrays. For fair comparisons, we apply the same total antenna power budget for all schemes and employ the power normalization of antenna area, as adopted in Theorem \ref{SNR_theorem}.  We fix array horizontal size as $M_x \Delta =  \lambda$ and therefore $M_x$ is set respectively as $2$ and $8$ for conventional arrays and HMIMOs.  By incorporating the impact of antenna area, it can be seen that given the same array aperture, even though HMIMO is equipped with more antennas, it still yields lower data rates than conventional arrays due to the reduced adjustable freedom in analog beamforming. This coincides with our analysis in Theorem \ref{SNR_theorem}.  To improve the beamforming quality of HMIMO, we explore the potential of a new feeding type, which is referred to as ``center feeding''. Specifically, unlike the common design where the signal  is fed from the left to the right  of the waveguide, we propose to feed the signal from the center of the waveguide to  both sides. This structure enables metamaterial antennas to sample the phase and amplitude of the reference wave in a symmetry pattern, enhancing the ability of coherent superposition  in the analog beamforming under limited precision. Note that our algorithm still works for center-feeding HMIMO after slightly modifying  $\mathbf{T}_i$. Fig. \ref{figure6} shows that with center feeding structures, the performance of HMIMO is significantly improved for both binary and grayscale constraints, thanks to the improved beamforming quality under simple circuits. Meanwhile, grayscale HMIMO almost achieves the same data rate as conventional arrays, which is very promising, especially considering the other benefits of HMIMO in terms of wiring layout, hardware complexity, cost, and power consumption.

\begin{figure}[t]
	\centering
	\includegraphics[width= 0.4\textwidth]{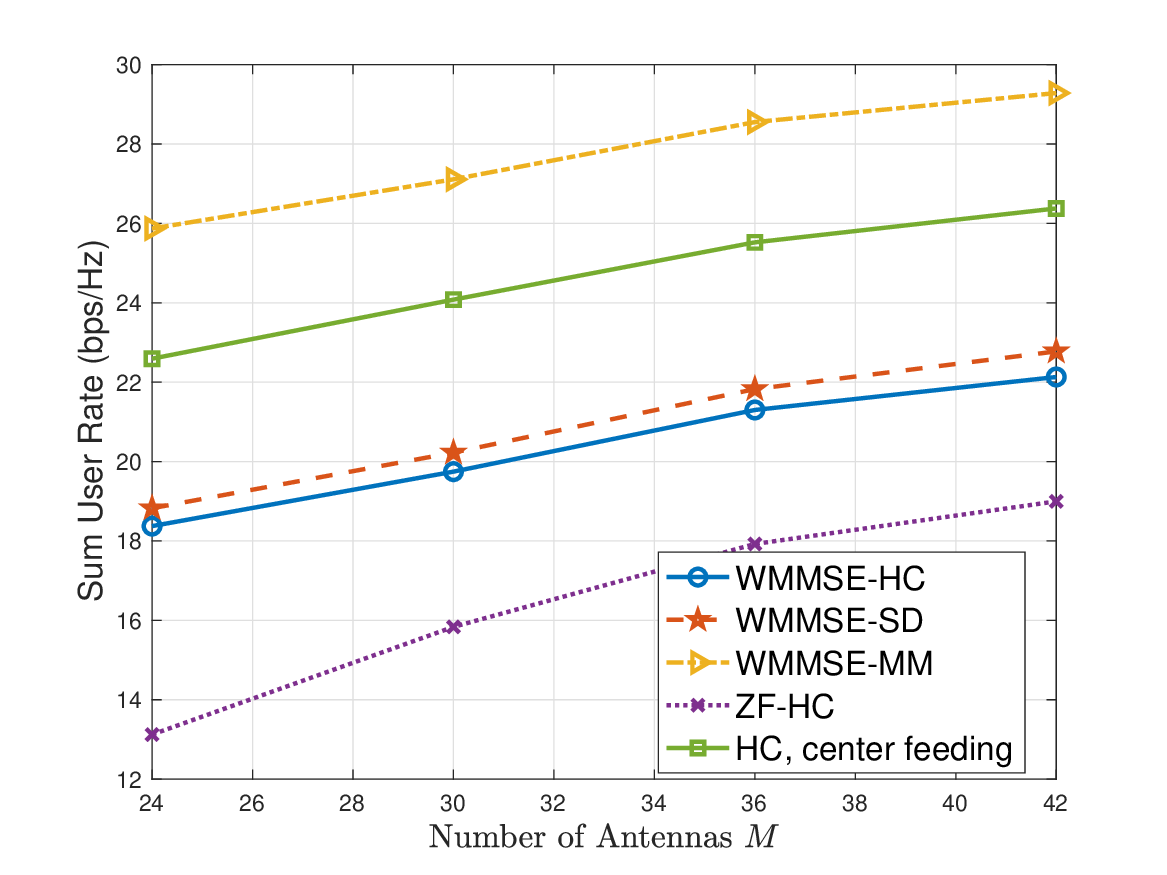}
	\DeclareGraphicsExtensions.
	\caption{Sum user rate versus the number of antennas, $ P_{i,m} = P_{\mathrm{tot}}/M $.}
	\label{figure5}
\end{figure}

  Fig. \ref{figure5} illustrates the performance of HMIMO with the number of antennas.  It is shown that the sum user rate is effectively increased with a larger number of antennas due to the enhanced beamforming gain.   It can be also observed     slight advantages of SD compared to HC, which again demonstrates the effectiveness of the HC algorithms.  Meanwhile, the performance improvement by using a center-feeding structure is again validated, demonstrating the great potential of this new structure.
  Besides, we compare our closed-form-based precoder design to the intra-cell ZF-based design. Advantages of our   algorithm can be seen. This is because our algorithm effectively tackles the intra-cell and inter-cell interference while the intra-cell ZF cannot mitigate the multicell interference.

\begin{figure}[t]
	\centering
	\includegraphics[width= 0.4\textwidth]{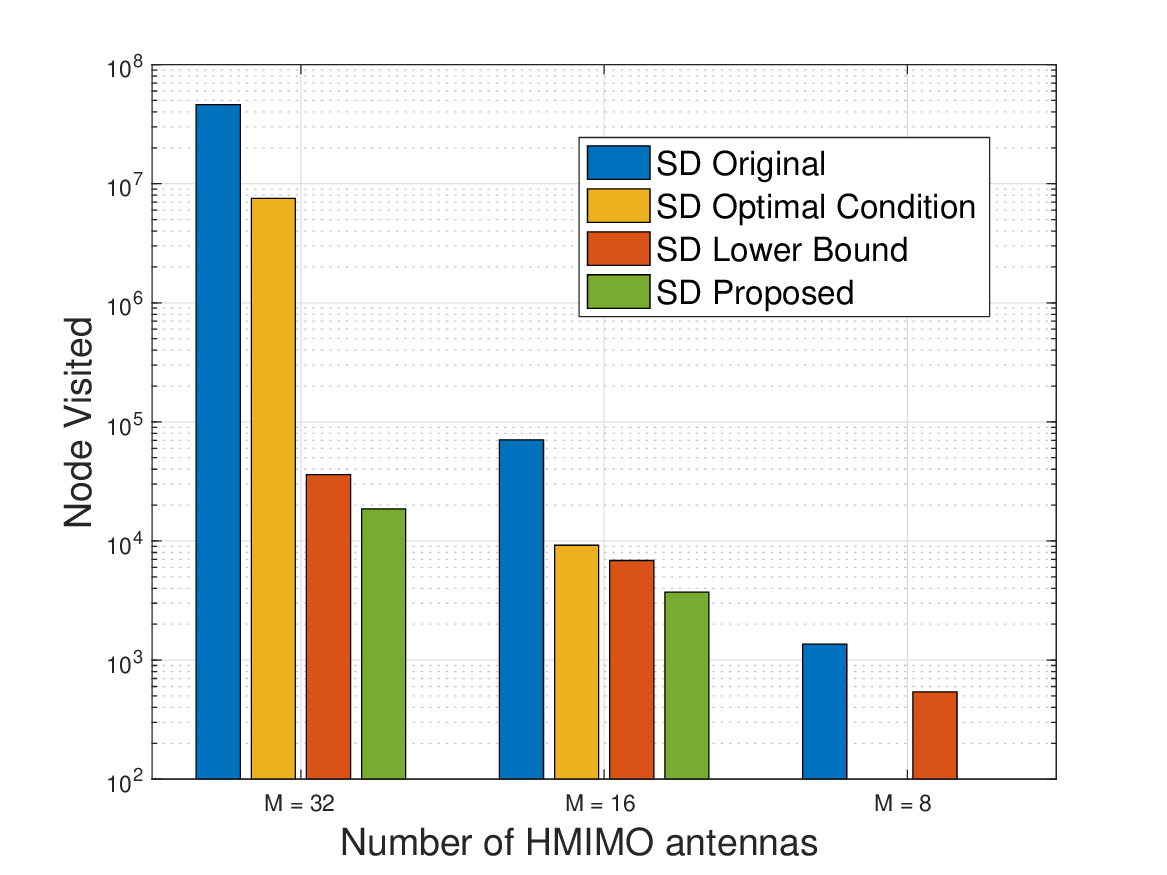}
	\DeclareGraphicsExtensions.
	\caption{Number of nodes visited in the search tree of SD.}
	\label{figure7}
\end{figure}

Next, we examine the effectiveness of the proposed accelerated SD algorithms in Fig. \ref{figure7}. It can be seen that the number of nodes visited in the SD algorithm increases with the number of antennas, due to the enlarged dimension of the optimization problems. Compared with the original SD algorithm, after exploiting the optimal condition (\ref{optimal_condition}) in the quadratic problem, the algorithm which is referred to as ``SD Optimal Condition'' can effectively reduce the number of nodes visited without additional computational complexity. When $M$ is small (e.g., $M=8$), even all nodes can be decided and thus zero complexity is needed. For the revised SD incorporating   tightened lower bounds (\ref{tighen_lowe_bound}), which is referred to as ``SD Lower Bound'', a huge number of nodes are pruned in the search space, although introducing some   additional complexity in each level.  Finally, it can be seen that the proposed accelerated SD algorithm which jointly exploits the optimal condition in the quadratic problem and the tightened lower bound, results in the lowest number of visited nodes.

\begin{table}[t]
	\centering 
	\caption{CPU Running Time for Binary HMIMO}	
	\begin{tabular}{|c|c|c|c|c|c|c}
		\hline
		& $M=16$ &  $M=24$  & $M=32$\\
		\hline
		WWSE Original SD&    0.8  &   58     &  $>7000$\\
		\hline
		WWSE  Proposed SD&    0.8      &   1   &   5.7   \\
		\hline
		WWSE-HC&      0.2   &   0.2     &  0.2\\
		\hline
		ZF-HC&  97    &      49     &            154        \\
		\hline
	\end{tabular}\label{tab2}
\end{table}

Finally, in Table \ref{tab2}, we showcase  the effectiveness of the proposed algorithm in terms of running time (in seconds). It can be seen that even though the original SD algorithm works well with small numbers of antennas, it leads to prohibitive-high ruining time for larger $M$. By contrast, the proposed SD algorithm can effectively reduce the running time. This demonstrates that although the introduced lower bound (\ref{tighen_lowe_bound})  needs more computations in each tree level, it effectively prunes a huge number of branches and, overall,   reduces the running time significantly. This indicates that it achieves a good trade-off between pruning  more branches      and  controlling additional computations. Meanwhile, it can be seen that the proposed HC algorithm spends very low computational time and thus is appealing to be applied in large-scale networks  with large numbers of antennas.  Furthermore, compared with ZF algorithms which are solved by CVX toolbox, our closed-form solution-based digital precoder design can be computed very quickly, which demonstrates its effectiveness.

\section{Conclusions} \label{section6}
This paper investigated HMIMO-enabled downlink multi-cell networks. We formulated the sum user rate maximization problem under the per-RF chain power constraint for  digital precoder and binary/Lorentzian phase constraints for HMIMO metamaterial elements. By exploiting the problem structure, we first designed the digital beamforming with closed-form solutions. Then, for binary-constrained HMIMO, we respectively proposed a sub-optimal HC algorithm and a global-optimal accelerated SD algorithm. For Lorentzian-constrained HMIMO, we proposed a low-complexity MM algorithm.  Moreover, we theoretically analyzed the SNR of HMIMO in a basic MISO scenario and explicitly revealed its advantages compared to conventional arrays. Numerical results demonstrated the promising performance of the proposed architectures.

\bibliographystyle{IEEEtran}
\bibliography{myref}

\end{document}